\documentclass[1.5p,review,12pt]{elsarticle}

\usepackage[T1]{fontenc}
\usepackage[utf8]{inputenc}
\usepackage[english]{babel}
\usepackage{graphicx}
\usepackage{wrapfig}
\usepackage[usenames,dvipsnames]{color}
\usepackage{amssymb}
\usepackage{multirow}
\usepackage{amsfonts,amsmath,amssymb}
\usepackage{epsfig}
\usepackage{subfig}
\usepackage{caption}
\usepackage{textcomp}
\usepackage[hang,flushmargin]{footmisc}
\usepackage{lineno}
\usepackage{mathrsfs}
\usepackage{amsthm}
\usepackage{amsmath}
\usepackage{bbm}
\usepackage{yfonts}
\usepackage[colorlinks=true]{hyperref}
\usepackage{algorithmic}
\usepackage[boxed]{algorithm2e}
\usepackage{mathrsfs}
\usepackage{xcolor}
\usepackage{comment}
\usepackage{csquotes}
\usepackage[free-standing-units,binary-units]{siunitx}
\usepackage{mathtools}
\usepackage{booktabs}
\usepackage[normalem]{ulem}
\usepackage[export]{adjustbox}
\usepackage{layout}

\usepackage[a4paper,
            left=0.7in,right=0.7in,top=1in,bottom=1in,
            footskip=.25in]{geometry}
            
\newcommand{\B}{\mathcal{B}}
\newcommand{\Bh}{\mathcal{B}^\mathrm{h}}
\newcommand{\Oe}{\Omega^{(\mathrm{e})}}

\newcommand{\Ge}{\Gamma^{(\mathrm{e})}}
\newcommand{\dB}{\partial\mathcal{B}}
\newcommand{\dBc}{\partial\mathcal{B}_\mathrm{C}}
\newcommand{\dBch}{\partial\mathcal{B}_\mathrm{C}^\mathrm{h}}
\newcommand{\dBcsl}{\partial\mathcal{B}_\mathrm{C,sl}}
\newcommand{\dBcst}{\partial\mathcal{B}_\mathrm{C,st}}
\newcommand{\gn}{g_\mathrm{n}}

\newcommand{\gt}{\mathbf{g}_\tau}
\newcommand{\gc}{g_\mathrm{c}}
\newcommand{\gp}{g_\mathrm{p}}
\newcommand{\gtd}{\dot{\mathbf{g}}_\tau}
\newcommand{\ur}{\mathbf{u}_\mathrm{r}}
\newcommand{\urd}{\dot{\mathbf{u}}_\mathrm{r}}

\newcommand{\tn}{p_\mathrm{n}}
\newcommand{\tc}{p_\mathrm{c}}
\newcommand{\ttt}{\mathbf{q}_\tau}
\newcommand{\en}{\varepsilon_\mathrm{n}}
\newcommand{\etd}{\dot{\varepsilon}_\tau}
\newcommand{\et}{\varepsilon_\tau}
\newcommand{\Dn}{\Delta_\mathrm{n}}
\newcommand{\Dz}{\Delta_0}

\newcommand{\Sr}{\mathcal{S}_\mathrm{r}}
\newcommand{\tf}{t_\mathrm{f}}
\newcommand{\np}{n_\mathrm{p}}
\newcommand{\ns}{n_\mathrm{s}}
\newcommand{\tttd}{\dot{\mathbf{q}}_\tau}
\newcommand{\Ac}{A_\mathrm{c}}

\DeclarePairedDelimiter{\norm}{\lVert}{\rVert}
\DeclareSIUnit\megapascal{\mega\pascal}

\journal{International Journal of Solids and Structures}

\graphicspath{{figures/}}

\begin{document}


\begin{frontmatter}
\title{A new finite element paradigm to solve contact problems with roughness}
\date{}

\author[1]{Jacopo Bonari\corref{cor1}}
\ead{jacopo.bonari@imtlucca.it}

\author[1]{Marco Paggi}
\ead{marco.paggi@imtlucca.it}

\author[2]{Daniele Dini}
\ead{d.dini@imperial.ac.uk}

\address[1]{IMT School for Advanced Studies Lucca, Piazza San Francesco 19, 55100 Lucca, Italy}
\address[2]{Department of Mechanical Engineering, Imperial College London, South Kensington Campus, London SW7 2AZ}

\cortext[cor1]{Corresponding author}

\begin{abstract}
This article's main scope is the presentation of a computational method for the simulation of contact problems within the finite element method involving complex and rough surfaces. The approach relies on the MPJR (eMbedded Profile for Joint Roughness) interface finite element proposed in [Paggi, Reinoso (2020) Mech Adv Mat Struct, 27:20 (2020)], which is nominally flat but can embed at the nodal level any arbitrary height to reconstruct the displacement field due to contact in the presence of roughness. Here, the formulation is generalized to handle 3D surface height fields and any arbitrary nonlinear interface constitutive relation, including friction and adhesion. The methodology is herein validated with BEM solutions for linear elastic contact problems. Then, a selection of nonlinear contact problems prohibitive to be simulated by BEM and by standard contact algorithms in FEM are detailed, to highlight the promising aspects of the proposed method for tribology.  
\end{abstract}


\end{frontmatter}

\begin{center}
\vspace{0.5cm}
\emph{Dedicated to Jim Barber's 80th birthday}
\end{center}

\section{Introduction}
During his long career, Professor James Barber has led many leading-edge advancements in the fields of continuum mechanics and contact mechanics. Since his dissertation~\cite{barber:1968}, he comprehensively exploited analytical methods to shed light on contact problems including friction~\cite{ahn:2008a,ahn:2008b,barber:2011a}, stability of thermo-elasticity~\cite{barber:1969,barber:1971,barber:1976}, surface roughness~\cite{barber:2003,barber:2013a,barber:2013b,paggi:2011c}. His research achievements have been recognized by highly cited publications and books~\cite{barber:2002,barber:2011b,barber:2018}.

Since the 1990s, the scientific problem of contact between rough surfaces, which was initially posed and investigated by mechanicians for tribological applications, has progressively attracted significant attention from researchers in other disciplines, especially physics and biology. Indeed, understanding how the multiscale features of surface roughness influence the overall emergent features of contact has fundamental implications for a wide range of technological and physical applications, see e.g. ~\cite{mueser:2017,vakis:2018,jacobs:2017,paggi:2020,goryacheva:2021,paggi:2022}. At the same time, the technological trend to engineer materials by tailoring their properties at the micro- and even at the nanoscales opens the issue of accurately representing all the relevant length scales for roughness and, at the same time, allows the simulation of nonlinear phenomena at the interface -e.g. friction or adhesion- and in the surrounding bulk -e.g. fracture, viscoelasticity, and plasticity.  

Research on this matter has seen significant progress since the 1950s, with the development of analytical and semi-analytical micromechanical contact theories departing from statistics of rough surfaces treated according to random process theory ~\cite{bowden:1950,archard:1957,greenwood:1966,greenwood:1967,bush:1975}. In the 1990s, the issue of resolution-dependency of contact predictions was raised with the advent of fractal models to synthetically represent roughness over multiple scales \cite{majumdar:1991,borribrunetto:1999,persson:2001}. 

This advancement paved the way for computational methods to simulate contact problems with roughness by directly including any given surface height field and avoiding assumptions on their statistical distributions. In this regard, the Boundary Element Method (BEM) (see ~\cite{andersson:1981,johnson:1985,keer:1999,bemporad:2015,xu:2019}) emerged as a powerful tool to analyze detailed 3D height fields, especially for frictionless and adhesionless contact problems and linear elastic materials.
This methodology has been proven to be computationally efficient since only the height field requires to be discretized and Green functions are used to simulate the response of the semi-infinite continuum. Attempts to generalize BEM to handle interface or material constitutive nonlinearities have been made within the last decades to include frictional effects \cite{paggi:2014,pohrt:2014,vollebregt:2014,anciaux:2010}, finite thickness of the domain~\cite{conway:1966,bentall:1968,greenwood:2012}, bulk viscoelasticity~\cite{carbone:2013,putignano:2014,putignano:2016}, interface adhesion \cite{carbone:2008,medina:2014,pastewka:2014,popov:2017,rey:2017}, wear~\cite{andersson:2011,brink:2021,frerot:2018}, plasticity~\cite{mayeur:1995,almqvist:2007,frerot:2019,frerot:2020}, lubrication~\cite{sahlin:2010}. However, such methodologies are difficult to be generalized to include all the above effects and some underlying assumptions cannot be lifted easily.



The Finite Element Method (FEM) would naturally allow gaining a deeper understanding of many key features of the subject which were once precluded with BEM, prime examples being the analysis of contact problems in finite elasticity, different nonlinear constitutive behaviors, and finite size geometries. However, the method comes with the cost of a remarkable increase in computational resources needed, together with the higher care necessary for a trustful discretization of the rough surface, avoiding artificial smoothing of fine scale geometrical characteristics of roughness. For these reasons, the use of FEM for the analysis of rough contacts has been limited to few studies regarding frictionless problems  compared to analytic solutions~\cite{hyun:2004}, plastic deformation~\cite{pei:2005}, finite strain indentation problems with Bezier-smoothed interface for the prediction of constitutive interface laws~\cite{bandeira:2004}, or studies devoted to the identification of the smallest  representative model size for micromechanical applications~\cite{yastrebov:2011,coutocarneiro:2020}.

In~\cite{paggi:2018}, the \emph{MPJR} approach has been introduced, which is capable of circumventing some of the criticalities stemming from the discretization of complex-shaped profiles according to FEM. The key idea consists in embedding the exact interface height field into a nominally smooth interface finite element, whose kinematics is borrowed from the Cohesive Zone Model ~\cite{ortiz:1999,paggi:2016}. Under the hypothesis of a rigid indenting profile, the exact deviation from planarity of the real geometry can then be restored by performing a suitable correction of the normal gap. This permits to model complex contacting geometries with simple low-order meshes, with a significant gain in the overall macroscopic geometry definition and contact solution algorithms. This regards two primary aspects: (i) the reduction of the high number of finite elements required for the explicit  discretization of the rough boundaries; (ii) the avoidance of corner cases caused by rapidly varying surface normal vectors that can induce a lack of convergence of contact search algorithms~\cite{wriggers:2006}.

The original MPJR formulation has been extended in \cite{bonari:2021a} to account also for friction in the partial slip regime. Moreover, it has been also employed to simulate ironing problems up to full slip and with finite sliding displacements for viscoelastic layers~\cite{bonari:2020}.

In the present article, the MPJR formulation is generalized in two different directions: $(i)$ 2D contact problems with rough profiles in the presence of friction and adhesive forces, as an example of a highly interface nonlinear problem; $(ii)$ 3D contact of rough surfaces with friction. The paper is structured as follows. In Sec.~\ref{sec:derivation}, the variational formulation of the interface finite element is detailed. In Sec.~\ref{sec:validation}, a set of numerical examples is presented to show the new capabilities of the approach. In Sec.~\ref{sec:conclusion} a summary of the results and an outlook of the future perspectives for tribological applications is provided.


\section{Variational formulation of contact problems with embedded roughness}\label{sec:derivation}
The framework detailed in the sequel regards the derivation of an interface finite element capable of simulating contact between a rigid surface $\Sr$ and a deformable bulk $\mathcal{B}$,
with a generic constitutive behavior, separated by a rough interface.  


\subsection{Contact with a conformal rigid surface}
The orientation of the boundary $\dB$ is determined by its outward pointing normal $\mathbf{n}$ and the kinematic quantities governing the contact problem. The normal gap, $\gn$, and the slip velocity, $\gtd$, are defined as:
\begin{align}
\gn &= \mathbf{n}\cdot(\ur-\mathbf{u}), & 
\gtd &= (\mathbf{I}-\mathbf{n}\otimes\mathbf{n})\cdot(\urd-\dot{\mathbf{u}}),
\label{eq:gap}
\end{align}
where $\ur$ and $\mathbf{u}$ represent, respectively, the displacements of the rigid surface $\Sr$ and of $\dBc$, which is the subset of $\dB$ where contact takes place.

The contact traction vector $\mathbf{t}$, related to the forces exerted by the contact of $\Sr$ over $\dBc$ can be expressed by means of the Cauchy theorem as $\mathbf{t}=\mathbf{T}\cdot\mathbf{n}$, where $\mathbf{T}$ is the Cauchy stress tensor. The split of $\mathbf{t}$ in its normal and tangential components, $\tn$ and $\ttt$ relative to $\dBc$, makes it possible to define the normal unilateral and tangential contact conditions.

If adhesive forces are neglected, the normal traction is always acting inward with respect to the boundary, and therefore it is negative. This allows us to summarize the conditions for normal contact in the set of relations known as Hertz-Signorini-Moreau (HSM) inequalities:
\begin{align}
\gn &\ge 0, & \tn &\le 0, & \gn \tn &= 0 & \text{on $\dBc$.}\label{eq:hsm}
\end{align}
Starting from this definition, a displacement-based normal contact constitutive relation can be defined by introducing a penalty parameter $\en$ which leads to the normal contact traction as: 
\begin{equation}
\tn = \en \gn.\label{eq:penalty}
\end{equation}

The introduction of a displacement-based traction law also permits to easily extend the analysis to adhesive problems via the definition of traction-penetration relations that regularize the HSM contact conditions. In this sense, the following constitutive relation is derived from a Lennard-Jones potential-like relationship in the normal direction~\cite{yu:2004,sauer:2009,mergel:2021} and reads:
\begin{equation}
\tn = \frac{A_H}{6\pi g_0^3}\biggl[
\biggl(\frac{g_0}{\gn}\biggl)^9-
\biggl(\frac{g_0}{\gn}\biggl)^3\biggl],\label{eq:LJ}
\end{equation}
where $A_H$ is the Amaker's constant characterizing the strength of adhesion and $g_0$ represents the equilibrium distance between two approaching half-spaces.

If the effect of friction is taken into account, then the contact response has to be differentiated depending on the status of the interface relative displacements in tangential direction. The contact domain is therefore given as:
\begin{align}
\dBc &=\dBcst \cup \dBcsl, \notag&
\dBcst \cap \dBcsl &= \varnothing.
\end{align}
In the equation above, the two subscripts \emph{st} and \emph{sl} denote the \emph{stick} and the \emph{slip} regions, respectively. The former is characterized by the absence of tangential relative motion between the bodies in contact, while the latter by a nonvanishing relative sliding which gives rise to tangential tractions opposing the relative movement. The solution of continuity in the contact subdomain boundary is a direct consequence of the non-linearity of the Coulomb law employed for modeling friction. 

\begin{figure}[b!]
\centering
\subfloat[][Hertz-Signorini-Moreau conditions vs. penalty approach.\label{fig:penalty}]
{\includegraphics[width=.4\textwidth]{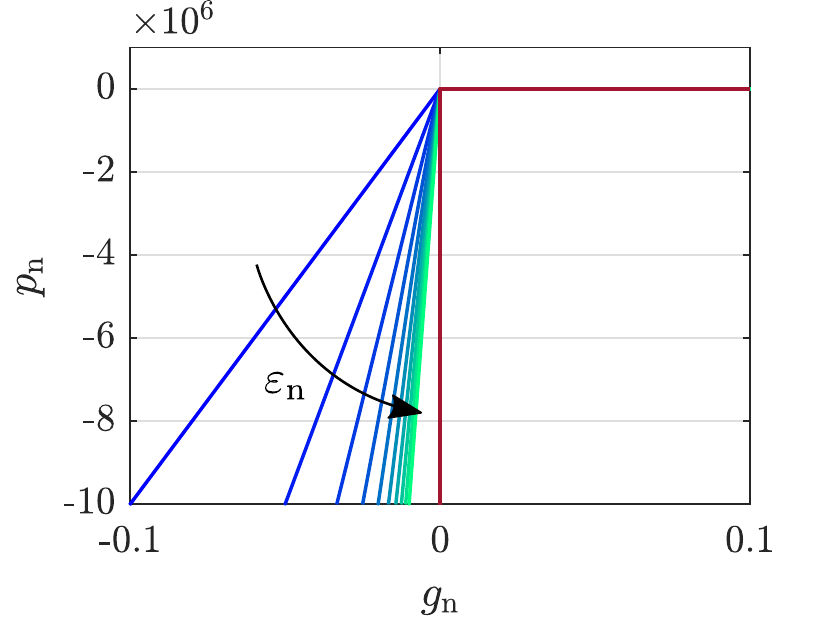}}\hspace{1mm}
\subfloat[][Coulomb friction law vs. regularised friction law.\label{fig:coulomb}]
{\includegraphics[width=.4\textwidth]{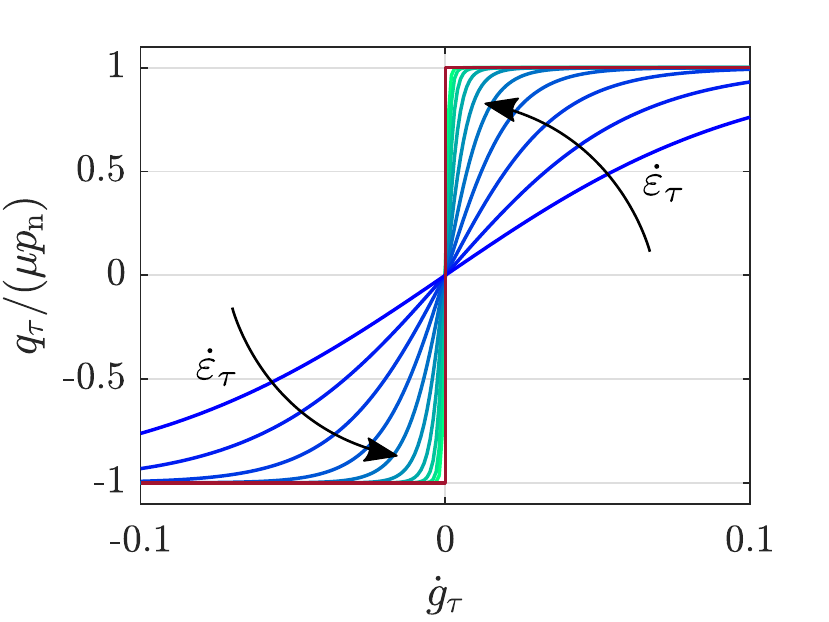}}
\caption{Normal and tangential constitutive relations for the traction field at the interface.}
\label{fig:reg_trac}
\end{figure} 

This can be expressed by the following set of equalities and inequalities:    
\begin{subequations}
\begin{align}
\gt &= 0,&\norm{\ttt} &\le \mu\lvert\tn\rvert &\text{on $\dBcst$}, \label{eq:coul} &\\
 \gtd &\ne 0, &\ttt &= \mu\lvert \tn\rvert\frac{\gtd}{\norm{\gtd}}\;\; &\text{on $\dBcsl$},
\end{align}
\end{subequations}
where $\gtd$ is the sliding velocity, and $\mu$ is the friction coefficient. According to Eq.~\eqref{eq:coul}, the tangential reaction can prevent relative sliding up to a limit value coincident with $\mu\lvert\tn\rvert$, above which relative sliding begins with a constant tangential reaction equivalent to the same threshold value. The interface behavior is depicted in Fig.~\ref{fig:coulomb}, together with the following regularized constitutive law employed for resolving the multi-valuedness in correspondence of the origin~\cite{feeny:1994}:
\begin{equation}
\ttt = \mu \lvert \tn \rvert \frac{\gtd}{\norm{\gtd}}\tanh{\frac{\norm{\gtd}}{\etd}}.
\label{eq:tanh}
\end{equation}
\textcolor{black}{The use of this specific regularization scheme is only a possibility amid different ones, see for example~\cite{simo:1992}. In the reference, the tangential response is modeled according to a Karush-Kuhn-Tucker (KKT) scheme for Coulomb friction, defined by the set of equations:}
\begin{subequations}
\begin{align}
\Phi = \norm{\ttt}-\mu\tn&\le0,& \gtd - \xi \frac{\partial \Phi}{\partial \ttt} & = 0, &\\
\xi &\ge 0, & \xi\Phi &=0,\label{eq:coulomb}
\end{align}
\end{subequations}
a regularisation can as well be defined on the slip rule, which after the introduction of a penalty parameter $\et$ reads:
\begin{equation}
\gtd - \xi \frac{\partial \Phi}{\partial \ttt} = \frac{1}{\et}\tttd.
\end{equation}

\textcolor{black}{The two different schemes deliver different errors introduced in the Coulomb friction law. On the side of Eq.~\eqref{eq:tanh}, the error stems from the lack of clear distinction between zones of stick and zones of slip, thus resulting in the introduction of a transition zone whose amplitude is strongly dependent on the chosen value of $\etd$. A clear and sharp distinction is only retrieved in the limit $\etd \to 0$. On the other hand, with the penalty regularization, the error is introduced as a difference between relative velocity and slip rate.}

\textcolor{black}{Each of the two possible choices comes with its own advantages and disadvantages, but they both provide robust constraints enforcement procedures. The use of Eq.~\eqref{eq:tanh} over the KKT penalized approach offers the advantage of directly linking tractions and displacements, with no need of defining trial stick and slip nodes, thus avoiding the necessity of setting up an additional loop for the identification of the correct stick and slip domains and the definition of a return map for the identification of the slip rate. This choice comes with the cost of having a less versatile implementation. The KKT formulation delivers exact results and the penalty regularization is just a possible way of proceeding. Stemming from the same KKT conditions, the problem can also be treated by exploiting lagrangian or augmented lagrangian schemes, with or without penalization. The same does not apply to Eq.~\eqref{eq:tanh}, being only a phenomenological interpretation of Coulomb's friction law. For the sake of completeness, the use of the hyperbolic tangent as regularizing function is only a possibility among different possible choices. Other functions that approximates tangential tractions arising from friction have been used and can be found in~\cite[Ch.\ 5, pp.\ 79--80]{feeny:1994,mostaghel:2005,pennestri:2016,vigue:2017,wriggers:2006}}

\begin{figure}[t!]
\centering
\includegraphics[width=0.4\textwidth,angle=0]{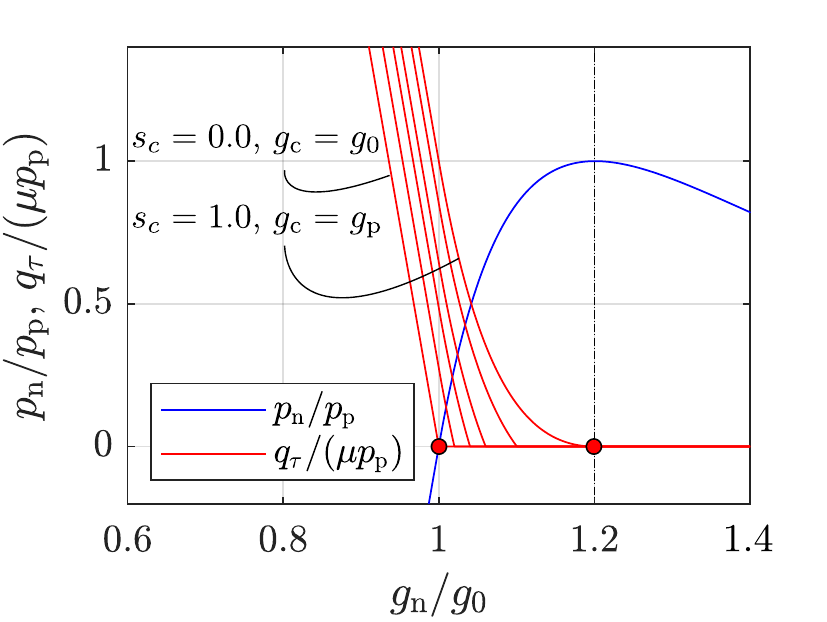}
\caption{Influence of cut-off normal gap $\gc$ over tangential tractions $q_\tau$.}\label{fig:LJ}
\end{figure}

When adhesion is also introduced, the tangential reaction expressed by Eq.~\eqref{eq:tanh} is modified as~\cite{mergel:2021}:
\begin{equation}
\ttt = \mu \bigl(\lvert \tn \rvert - \tc \bigl)H\bigl(\gc-\gn\bigl) \frac{\gtd}{\norm{\gtd}}\tanh{\frac{\norm{\gtd}}{\etd}},
\label{eq:tanh_adh}
\end{equation}
where $p_\mathrm{c}$ is the value of the normal traction corresponding to a specific cut-off normal gap $\gc$, and $H(x)$ is the Heavyside step function. In this way, the effect of the adhesive tractions on the frictional forces can be modulated. \textcolor{black}{Introducing $H(x)$ in Eq.~\eqref{eq:tanh_adh} makes the tangential tractions field only $\mathcal{C}^0$ differentiable, unless the condition $\gc = \gp$ is met, being $\gp$ the normal gap related to the pull-out normal traction. Since the global (and unique) point of maximum for the normal tractions is located in correspondence with this point, this is the only value for which $\ttt$ could reach a null value smoothly, Fig.~\ref{fig:LJ}. On the other hand, imposing $\gc = g_0$, the classic Coulomb law can be retrieved, in the sense that no tangential forces are present for positive normal tractions. In this latter case, the system's full slip state can be more easily assessed since a perfect correspondence between tangential and normal tractions scaled by $\mu$ is guaranteed.}

The contribution of the interface to the weak form of the boundary value problem can be written by means of the virtual work principle as:
\begin{equation}
\delta\boldsymbol{\Pi}=
\int_{\dBc} (\tn \cdot \delta \gn + \ttt \cdot \delta \gt)\,\mathrm{d}s.
\label{eq:finel}
\end{equation}

\begin{figure}[b!]
\centering
\includegraphics[width=0.5\textwidth,angle=0]{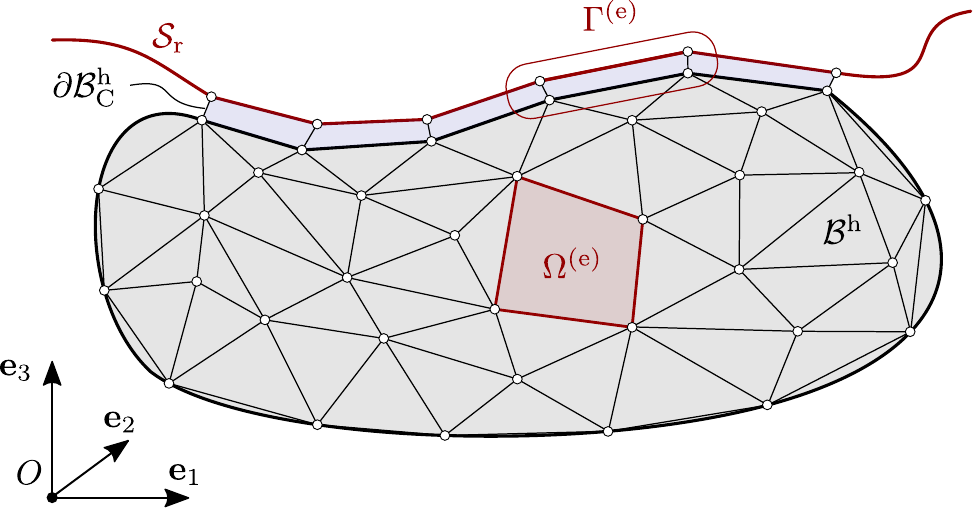}
\caption{FEM approximation of the bulk and the interface.}\label{fig:approx}
\end{figure}

The solution of the contact problem in a finite element framework requires the geometrical approximation of $\B$ and of the contacting interface $\dBc$, an operation that paves the way for their discretization into finite elements. The process can be formalized as:
\begin{align}
\B \approx \Bh&=\bigcup\limits_{\mathrm{e}=1}^{n_\Omega}\Oe, &
\dBc \approx \dBch &= \bigcup\limits_{\mathrm{e}=1}^{n_\Gamma}\Ge,
\end{align}
where $\Oe$ represents a single finite element composing the geometric approximation $\Bh$ of the bulk $\B$, while $\Ge$ describes the discretization of $\dBch$, in its turn approximation of $\dBc$, Fig.~\ref{fig:approx}.

\begin{figure}[h!]
\centering
\subfloat[][\emph{quad} finite element.\label{fig:elsa}]
{\includegraphics[width=.3\textwidth]{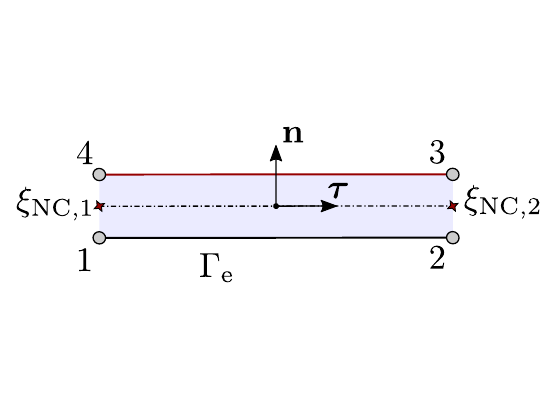}}\hspace{5mm}
\subfloat[][\emph{hex} finite element.\label{fig:elsb}]
{\includegraphics[width=.3\textwidth]{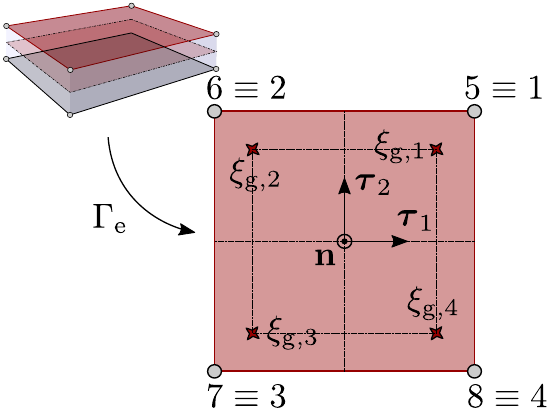}}
\caption{$2D$ and $3D$ interface finite elements.}
\label{fig:els}
\end{figure}

Given the hypotheses of conformal contact interface, matching nodes on the overlying surface can be identified in correspondence to the ones on the bulk's boundary, and $\Ge$ can be defined as an interface finite element in analogy to CZM for fracture~\cite{ortiz:1999}. Here, they are characterized by two facets, one belonging to $\dBch$ and one to the contacting rigid surface; the relative displacement of a couple of matching nodes is responsible for the exchange of reaction forces across the interface thanks to the defined constitutive relations. Figure~\ref{fig:els} shows their layout for a 2D case, where the element coincides with a collapsed four nodes quadrilateral (\emph{quad}), and in 3D, where the element is analogous to a collapsed eight nodes hexahedral (\emph{hex}).

\subsection{Gap field correction to account for roughness}
The basic characteristics of the interface finite element derived above are suitable for the solution of conformal contact problems under small displacements assumptions, \textcolor{black}{with characteristics analogous to a segment-to-segment approach with fixed pairings. It has to be remarked that up to this point the formulation is also valid for the solution of deformable-to-deformable contact, since the only requirement to be respected is the presence of a conformal interface.}

In~\cite{paggi:2018,bonari:2021}, an extension has been proposed to analyze rigid to deformable non-conformal contact problems, from standard curved indenters up to quasi-fractal wavy or fractal rough surfaces. While the interested reader is referred to the articles above for a detailed derivation of the method, in the following only the underlying idea is presented. 

\begin{figure}[b!]
\centering
\subfloat[][Actual rough contact.\label{fig:rough_flat_a}]
{\includegraphics[width=.4\textwidth]{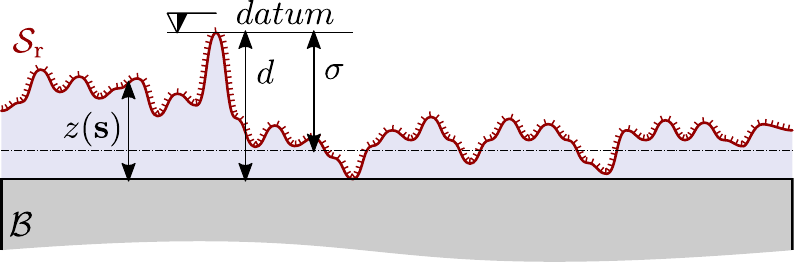}}\hspace{5mm}
\subfloat[][Equivalent FE interface.\label{fig:rough_flat_b}]
{\includegraphics[width=.4\textwidth]{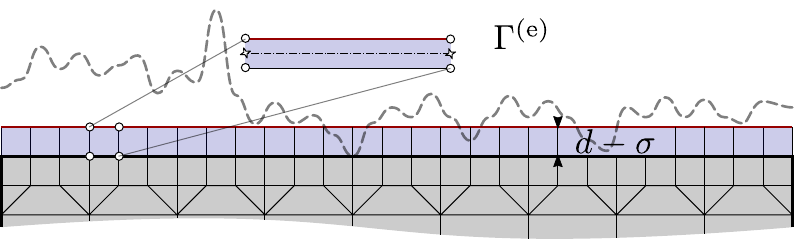}}
\caption{Interface discretization with embedded roughness.}
\label{fig:rough_flat}
\end{figure}

According to Fig.~\ref{fig:rough_flat}, starting from the conformal configuration, a \textcolor{black}{rigid} contacting surface of arbitrary geometry can be taken into account thanks to a suitable correction of the gap field defined in Eq.~\eqref{eq:gap}. If a local reference system is set in correspondence of $\dBch$, an elevation field marking the deviation from planarity between the smoothed and the real geometry can be introduced. In the simplest case of an interface geometry analytically defined as a function $z(\mathbf{x})$, the corrected gap reads $\gn^\ast = \gn + z(\mathbf{x})$. The use of the modified gap in the derivation of the system's stiffness matrix allows accounting for the complex geometry without the need to actually consider it explicitly during the FE discretization process.

\textcolor{black}{Once the correction of the gap function is introduced, the method applied to two deformable bodies is still able to account for the effect of the elastic contact interactions in the bulk. However, second order effects, which would modify the local elevations of the embedded rigid profile, are not accounted for at the moment. A possible strategy to overcome this aspect could be the introduction of an update of the embedded elevation function $z(\mathbf{x})$ based on the deformation of the underlying bulk.}


\textcolor{black}{Therefore, considering a rigid indenter, the contact problem can be simulated with a standard FE discretization of the bulk material, accompanied by a single layer of interface finite elements in correspondence of the active set of contact, which stores the contact geometry information in the form of a corrected gap.}

\begin{figure}[t!]
\centering
\subfloat[][The rigid indenting surface is held fixed, BCs are applied on the deformable part of the mesh only.\label{fig:bcsa}]
{\includegraphics[width=.3\textwidth]{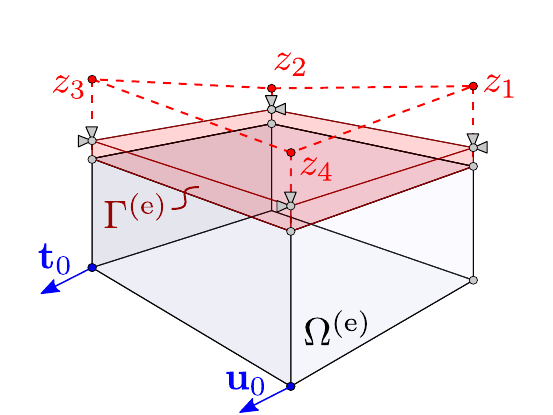}}
\subfloat[][A fictitious rigid bulk is linked to the rigid indenting surface via the upper nodal pairs of the interface finite elements. BCs can be applied both to deformable and rigid indenting surfaces.\label{fig:bcsb}]
{\includegraphics[width=.3\textwidth]{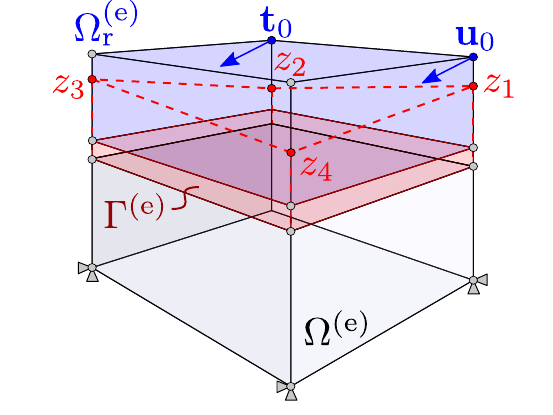}}
\subfloat[][Dirichlet BCs can be directly applied to the rigid surface as a prescribed act of motion.\label{fig:bcsc}]
{\includegraphics[width=.3\textwidth]{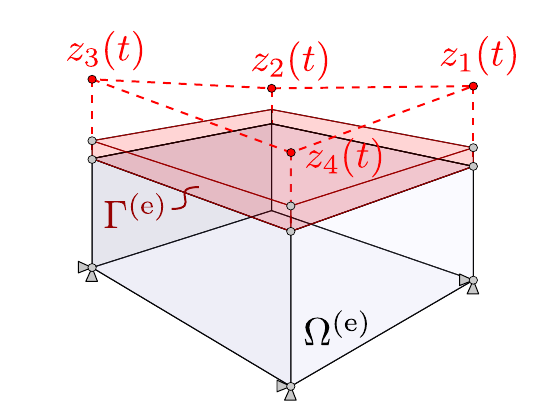}}
\caption{Different procedures for enforcing Dirichlet ($\mathbf{u}_0$) or Neumann ($\mathbf{t}_0$) BCs over the rigid indenter and the deformable bulk.}
\label{fig:bcs}
\end{figure}

\textcolor{black}{At this stage, the application of boundary conditions (BCs) to the model can be performed by constraining the nodal pair of the interface finite elements opposite to the bulk and applying load in the form of Dirichlet or Neumann BCs to the bulk nodes, i.e. considering the surface to be fixed with motion only possible for the deformable body, Fig.~\ref{fig:bcsa}. An option for the application of load in the form of a rigid act of motion or concentrated or distributed forces directly to the indenter is possible with the deployment of an additional layer of standard finite elements on the free side of the interface, and apply them the desired BCs, Fig.~\ref{fig:bcsb}. For preserving the hypothesis of rigidity, however, a high level of stiffness compared to the bulk's material has to be assigned to them, where $\mathbf{t}_0$ and $\mathbf{u}_0$ represent applied nodal forces and displacements, respectively.}

\textcolor{black}{A third approach can be conceptualized as well, where a rigid act of motion is directly applied to the rigid surface in the form of suitable time dependence of the elevation field, that in this case would read $z=z[\mathbf{x}+\boldsymbol{\Omega}(t)]$, where $\boldsymbol{\Omega}(t)$ is a three dimensional curve, parametrized in time, that describes the act of motion of the rigid surface, Fig.~\ref{fig:bcsc}. The study of this methodology of constraint enforcement goes beyond the scope of the present publication and is left for further studies. Some preliminary result, though, has been presented in~\cite{bonari:2020}, where the concept has proven to be applicable in the context of the analyses of tangential motion over long slipping distances, nevertheless still in the context of a small strain theory. It has to be remarked that the ability to consider long slipping distances is actually a limitation of the implementation proper to the first two ways of BCs enforcement presented in the article. Given that, in compliance with a contact scheme that requires matching nodes at the interface, the variation of the elevation field $z(\mathbf{x})$ consequent to lateral sliding is not taken into account, therefore limiting the analysis to infinitesimal sliding distances.}  

\textcolor{black}{In conclusion of this section,} two different approaches \textcolor{black}{are presented} for the assignment of the correct elevation field to each elements' Gau\ss~points. The rough surfaces employed in the contact simulation can be either hard-coded in the element routine (in the case it can be defined analytically) or stored in an external file as a three columns matrix of $[x,y,z]$ values and prompted as a look-up table (this latter solution being necessary in the case the surface to be used directly comes from topographic measurements, such as those obtained from photogrammetry or confocal profilometry).

\section{Numerical examples}\label{sec:validation}
In this section, new results related to the 2D analysis of rough profiles and verification against BEM simulations are provided.
Then, adhesive contact problems, also including friction, are solved for wavy profiles. Moreover, \textcolor{black}{one benchmark test and two bigger scale applications} are shown to prove the capability of the method to handle full scale 3D complex morphologies.

\subsection{MPJR validation in 2D with BEM}\label{sec:RMD}
The normal frictionless indentation problem \textcolor{black}{of an elastic layer of finite depth by a rough profile} is herein addressed, and the results compared with the BEM solution related to the same problem. The profile is obtained using a Random Midpoint Displacement (RMD) algorithm  often employed for the generation of rough surfaces characterized by a given fractal dimension $D$~\cite{mandelbrot:1977}, see also \cite[Ch.\ 16, pp.\ 357--358]{paggi:2011c,fractals:1988,barber:2018} for more details, \textcolor{black}{and~\cite{perez:2019} for a possible numerical implementation of this fractal surface generation algorithm, capable of creating elevation fields with given Hurst exponent $H$ and height probability distribution.}

\textcolor{black}{The 2D profile used has been obtained as the section cut of a 3D rough surface generated exploiting the numerical procedure exposed in~\cite{paggi:2011c}. The section cut is performed in correspondence with its highest summit, i.e. the first point supposed to come into contact during the indentation process.} In the benchmark test we set a surface fractal dimension $D=2.2$, a random seed uniformly distributed in $[-1,+1]$ and a random function with Gaussian distribution and initial standard deviation $\sigma_0 = 2.357$ to generate a height field spanning over one decade of length scales, thus characterized by $N = 2049$ elevation points equally spaced in the horizontal direction.

\begin{figure}[b!]
\centering
\includegraphics[width=0.4\textwidth]{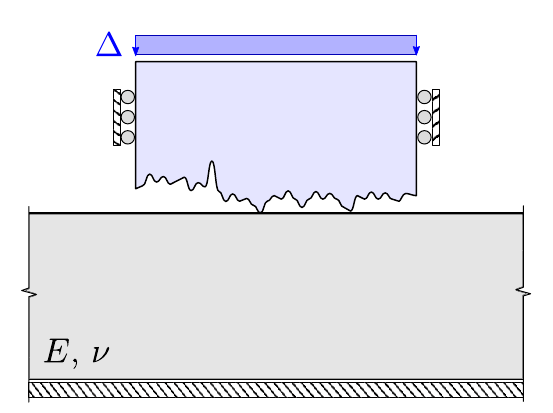}
\caption{Sketch of the problem under examination.}
\label{fig:wedge}
\end{figure}

The profile is considered as the boundary of a rigid \textcolor{black}{indenter} that makes contact with a linear elastic layer of finite unitary depth $b$ that spans indefinitely in the horizontal direction and rests on a frictionless rigid foundation. The rough profile spans horizontally over a length of $2b$ and has an overall height of $g_0 = 1.0\times10^{-2}b$, measured from the lowest valley to the highest peak.
The elastic layer is characterized  by Young's modulus $E=1\,\si{\mega\pascal}$ and Poisson's ration $\nu=0.3$.
The load is applied under displacement controlled conditions. A downward imposed vertical displacement linearly increasing from zero up to the value of $\Delta_0 = 3g_0$ is applied in fifteen pseudo-time steps. The problem addressed is sketched in Fig.~\ref{fig:wedge}, plane strain assumptions hold.

\begin{figure}[h!]
\centering
\subfloat[][Bulk dimensions and discretization.\label{fig:msha}]
{\includegraphics[width=.35\textwidth]{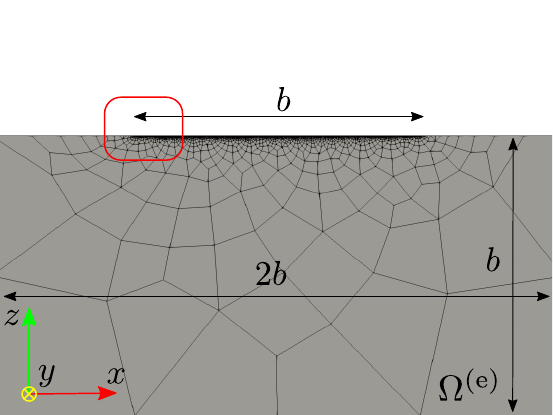}}\hspace{5mm}
\subfloat[][Close up of the edge of the contacting interface.\label{fig:mshb}]
{\includegraphics[width=.35\textwidth]{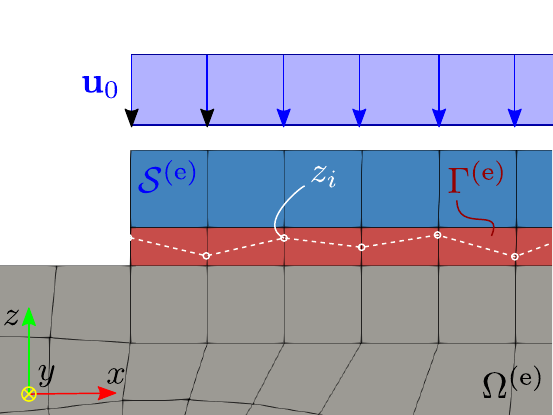}}
\caption{FEM implementation required for the problem's solution.}
\end{figure}

For its solution employing the proposed method, the following FEM implementation has been set up. First, the elastic layer has been modeled using standard \emph{quad} bilinear finite elements. Since the solution is focused on the contact interface, grading has been performed resulting in a finer resolution in the zone of interest, where a one-to-one correspondence holds between the interface nodes and the profile sampling points, Fig.~\ref{fig:msha}, finally, the bulk is truncated in the horizontal direction after a distance of $b/2$ on the left and right sides of the contact zone, since after mesh convergence studies a higher length has proven not to affect the quality of the results.

In the contact zone, a single layer of interface finite elements $\Ge$ is deployed over the bulk elements, their lower nodes matching the  boundary nodes of the bulk, for a total of $n_\Gamma = N-1$. This is where the geometric pieces of information of the rough profile are stored elementwise, and the actual normal gap is evaluated as a correction of the original one, Fig.~\ref{fig:mshb}. The arrangement is completed by a single structured layer of standard \emph{quad} elements, tied with the interface finite elements, much stiffer than the bulk's element, devoted to receiving the enforcement of the boundary conditions and transmitting them to the upper nodal pair of the interface finite element, cfr. Fig.~\ref{fig:mshb} and~\ref{fig:bcsc}. In the specific, they have been assigned a Young's modulus $E_\mathrm{r} = 1.0\times10^{3}E$. Finally, a normal penalty parameter $\en = 1.0\times10^{3}E/b$ has been used.

For providing a benchmark solution, the same problem has been solved by exploiting a BEM framework developed for $2D$ plane strain contact problems. In the specific, the Green function employed reproduces the displacement field occurring at the free boundary of a linear elastic layer of finite depth resting frictionless on a rigid foundation, when uniform pressure is applied over a limited strip. Its expression can be found in~\cite{bentall:1968}. With the only difference of Green functions employed, the remaining BEM implementation and related details are the same used in~\cite{bemporad:2015}.

\begin{figure}[b!]
\centering
\includegraphics[width=0.6\textwidth]{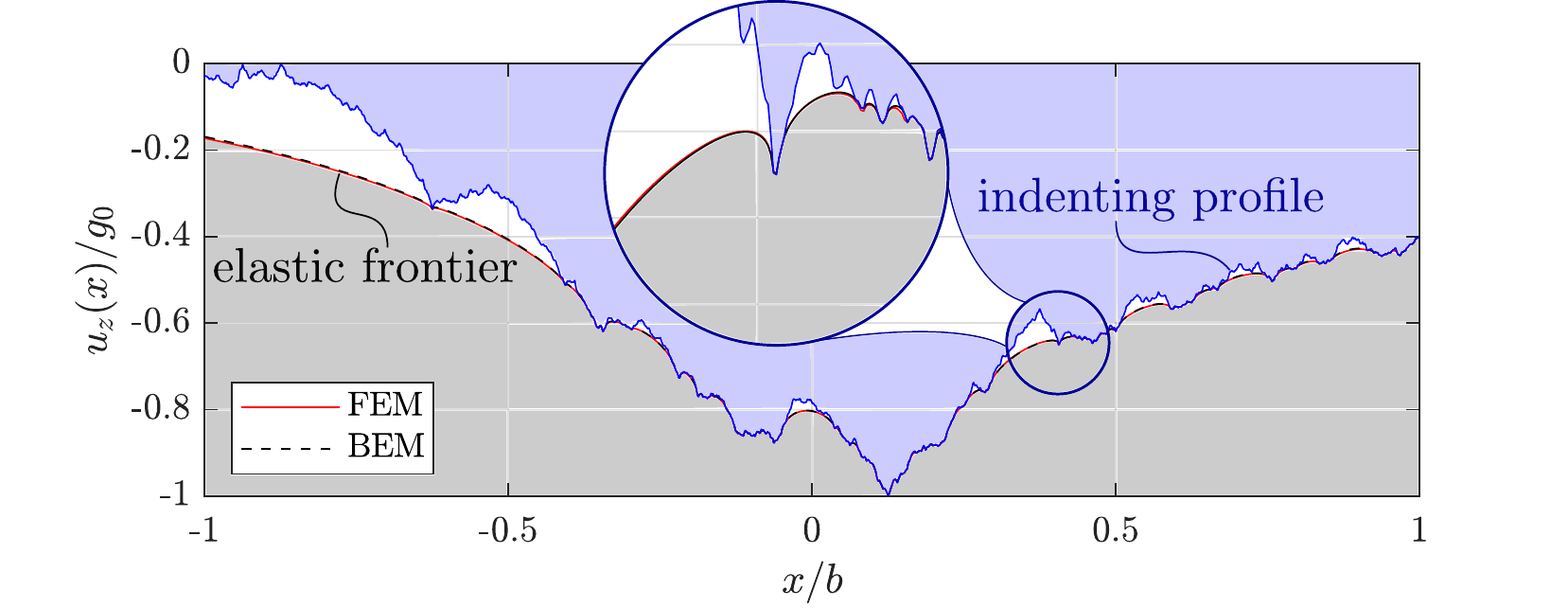}
\caption{Deformation of the elastic frontier under imposed normal far field displacement and detailed indenting profile geometry.}
\label{fig:BEM2048_disp}
\end{figure}

The shape of the indenting profile can be appreciated in Fig.~\ref{fig:BEM2048_disp} (solid blue line), together with the qualitative solutions delivered by the FEM (solid red line) and BEM (black dashed lines) procedures, in terms of surface displacements $u_z(x)$. The presented plot is a snapshot taken for $t=\tf/3$ so that the imposed displacement corresponds to $g_0$. Qualitatively, a perfect agreement is observed between the two solutions.

\begin{figure}[h!]
\centering
\includegraphics[width=0.6\textwidth]{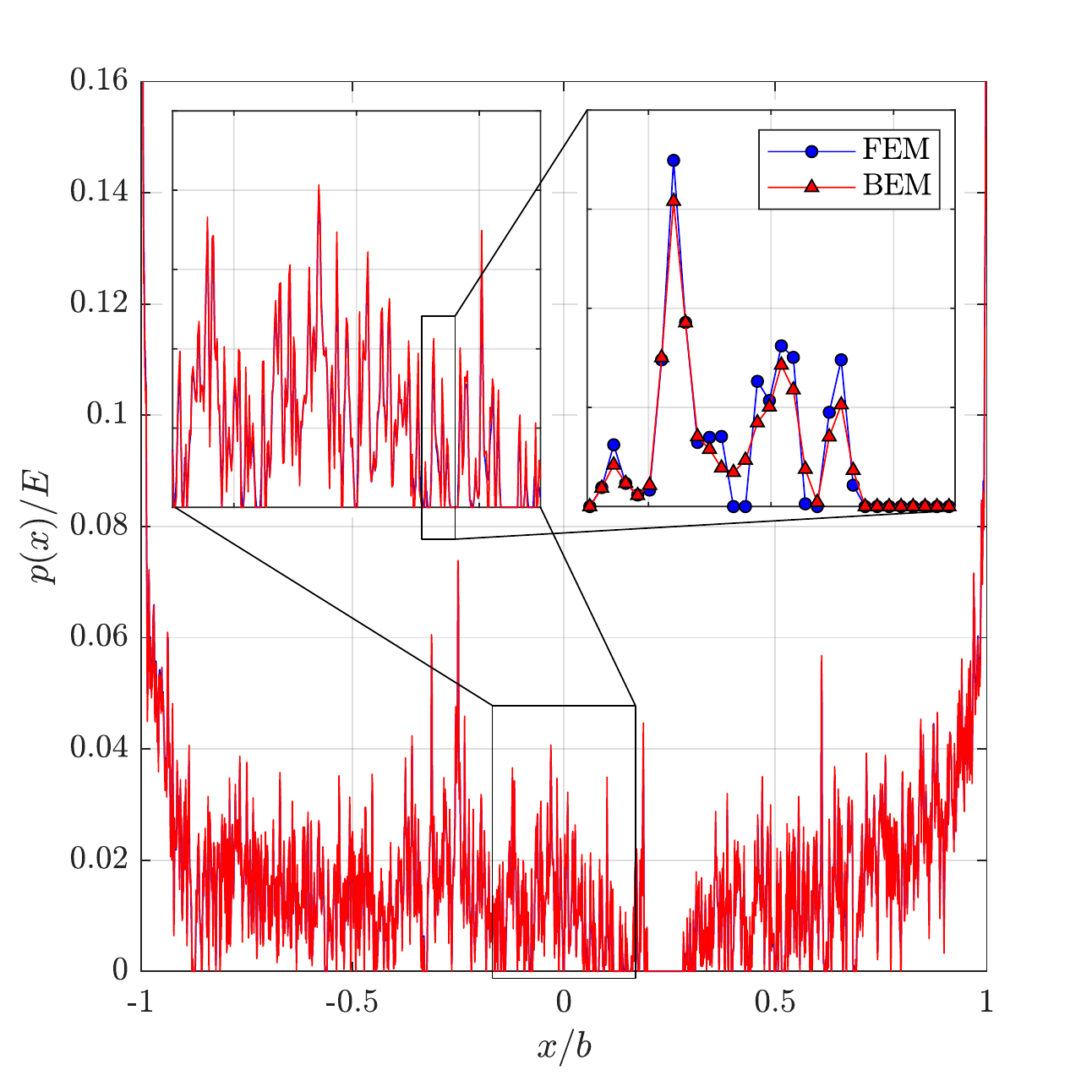}
\caption{Comparison of BEM vs. FEM. Results in terms of normal tractions field at the interface.}
\label{fig:fbstre}
\end{figure}

\begin{figure}[h!]
\centering
\includegraphics[width=0.4\textwidth]{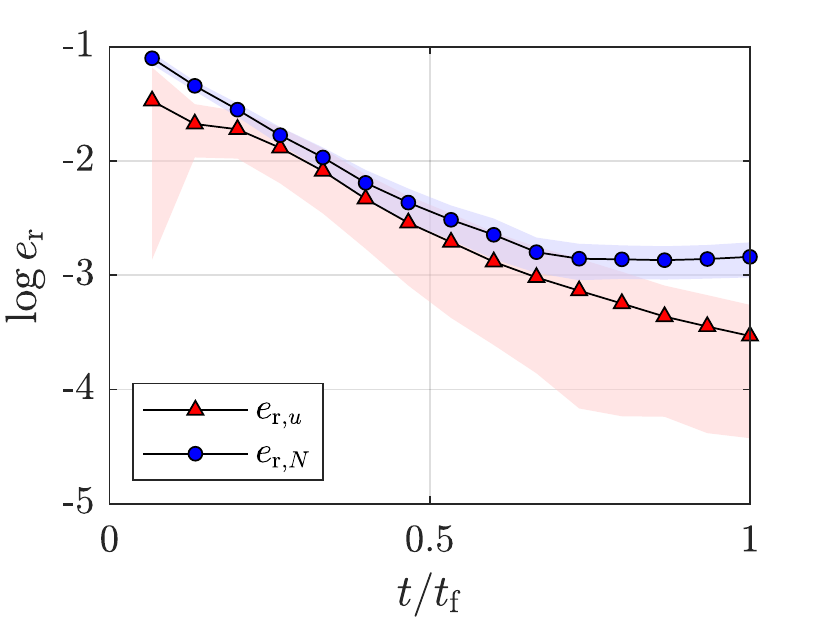}
\caption{Error estimate.}
\label{fig:error}
\end{figure}

A quantitative comparison is now made in terms of interface normal tractions $p(x)$. For pursuing statistically representative results, different simulations have been performed. Given the same loading conditions, mesh sizes, and mechanical parameters, different profiles to be tested are generated. More specifically, ten different values of the Hurst exponent have been set, linearly varying from $H=0.75$ to $H=0.85$. For each of these values, ten different random seeds have been used in the generation process, for a total of $100$ different profiles. Figure~\ref{fig:fbstre} shows a specific solution, related to the normal traction field along with the contacting interface, for both FEM (blue round markers) and BEM solutions (red triangular markers), at a given time step. In the top-right magnified panel, some small discrepancies in the two results can be noticed, but still, very good accordance can be appreciated. In the authors' opinion, such differences are to be ascribed to the kind of profile employed here, i.e. a scattered elevation field which could be considered a worst case scenario in the context of a contact mechanics problem solved using FEM. This hypothesis is supported by the perfect agreement that, on the other hand, can be appreciated if a benchmark on contact tractions is performed for what concerns a smooth indenting profile, see for example~\cite{bonari:2021}. Finally, Fig.~\ref{fig:error} quantitatively reports the mean absolute relative error in terms of displacement at the interface and total reaction force between FEM and BEM, evaluated over all the profiles employed, for every point of the contacting interface, plotted for every time step of the analysis. The transparency bands denote the variation of the standard deviation of the error distribution for every time step. The expression of the error reads:
\begin{equation}
e_\mathrm{r}=\frac{1}{\ns\np}\sum_{i=1}^{\ns}\sum_{j=1}^{\np}\biggl\lvert\frac{u^{(\mathrm{f})}_{ij}-u^{(\mathrm{b})}_{ij}}{u^{(\mathrm{f})}_{ij}}\biggl\lvert,
\end{equation}
where superscripts $(\mathrm{f})$ and $(\mathrm{b})$ stand for BEM and FEM simulations, respectively, where an analogous expression can be drawn for the error over normal reaction forces $N$. Both the error estimates deliver very good results with values rapidly approaching zero as the load increases.

\pagebreak

\subsection{Frictional response with adhesion for a Wavy profile}

The second example is characterized by more complex constitutive relations inclusive of friction and adhesion. The adopted profile is herein simpler, but it is still comprehensive of the standard difficulties characterizing the solution of such types of problems using other state-of-the-art numerical methods, namely the non-compactness of the contacting domain, the use of non-convex constitutive relationships, and the presence of finite bulk dimensions.

The numerical simulation consists of the indentation problem of a finite depth elastic layer by a rigid wavy profile made of the superposition of two harmonics, deriving from the truncation of a Weierstrass profile defined by the following expression:
\begin{equation}
z(x)=g_0\sum_{i=0}^{\infty} \gamma^{(D-2)i}\cos\left(2\pi\dfrac{\gamma^i x}{\lambda_0}\right).
\end{equation}
Its geometry is obtained by setting $H=0.75$, $\gamma = 5$, $z_0=1.0\times 10^{-1} l_0$ and $\lambda_0 = 2l_0$, where $H$ and $\gamma$ are the Hurst exponent and the base of the wavelength's geometric progression across the scales; $\lambda_0$ and $z_0$ are the fundamental wavelength and amplitude.
The bulk has been modeled as a rectangular elastic block. It is considered perfectly bonded in correspondence with the lower base, and periodic boundary conditions have been applied on both the vertical sides, in correspondence of $x = \pm l_0 = \pm 10\,\si{\micro\metre}$. A Young's modulus of $E = 20.0\,\si{\megapascal}$ and a Poisson's ratio $\nu=0.3$ have been assigned to the linear elastic bulk.
The model employed for reproducing the tangential behavior is in accordance with Eq.~\eqref{eq:tanh_adh}, with a coefficient of friction $\mu=0.2$ and a cut-off on friction forces $\gc = \g_0$.
The two parameters chosen for modelling the adhesion law are the max adhesive pressure $p_\mathrm{m} = 0.330\,\si{\megapascal}$ and a work of adhesion $W = 0.027\,\si{\joule/\metre^2}$. The chosen values result in $\gp \approx 1.0\times 10^{-2} l_0$, thus keeping the transition from negative to positive normal contact tractions appreciable employing a reasonable fine discretization for the interface. In the specific, for such a case, $2048$ interface finite elements have been employed for sampling a region corresponding to the fundamental wavelength $\lambda_0$.

The finite element arrangement is analogous to the one presented in the previous case study. Standard finite elements have been used for modeling the bulk, a single layer of interface finite elements is deployed over the active contact zone and on top of that a layer of standard finite elements, much stiffer than the bulk elements, is devoted to the application of BCs.
 
\begin{figure}[b!]
\centering
\subfloat[][Normal loading phase.\label{fig:pzwma}]
{\includegraphics[width=.5\textwidth]{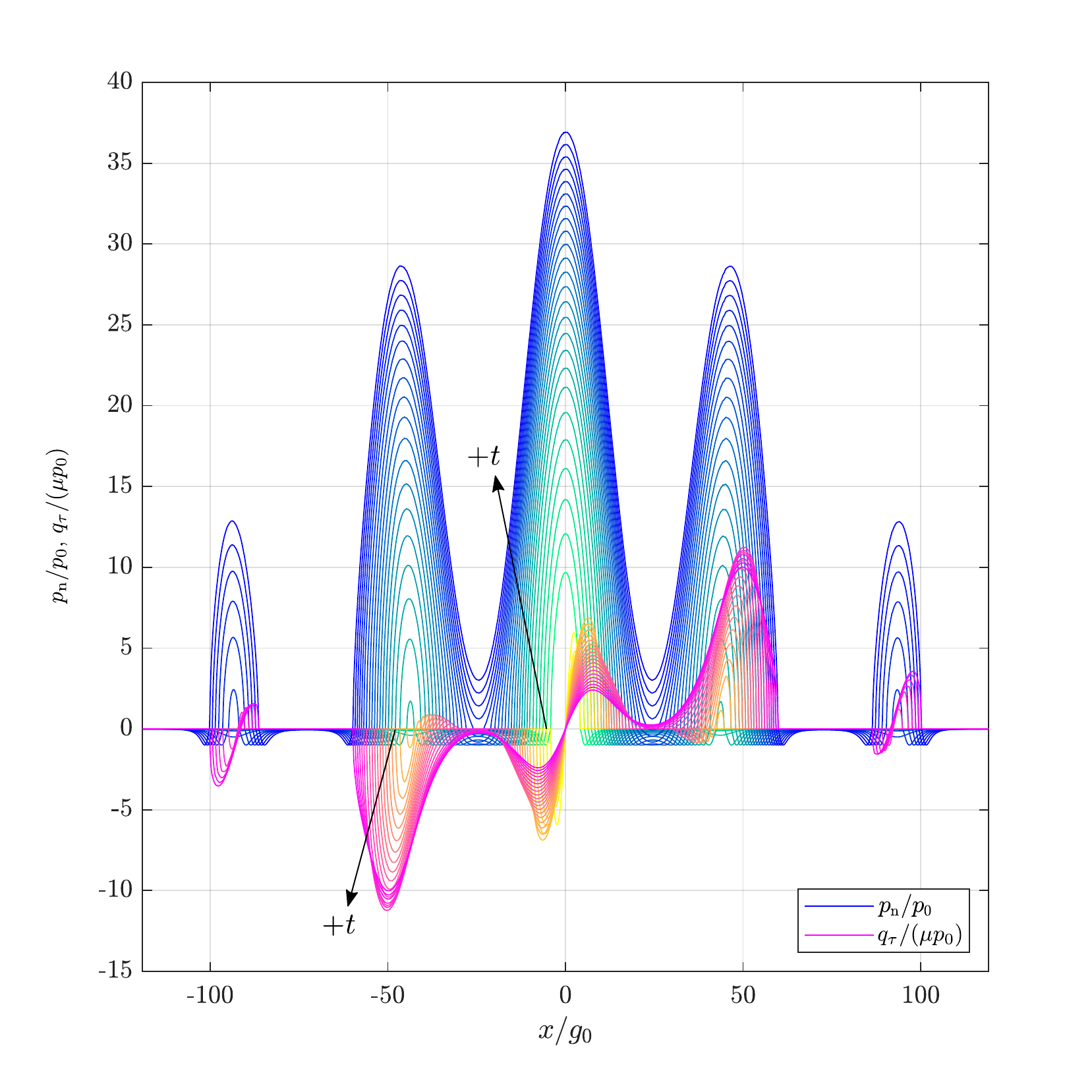}}
\subfloat[][Tangential loading phase.\label{fig:pzwmb}]
{\includegraphics[width=.5\textwidth]{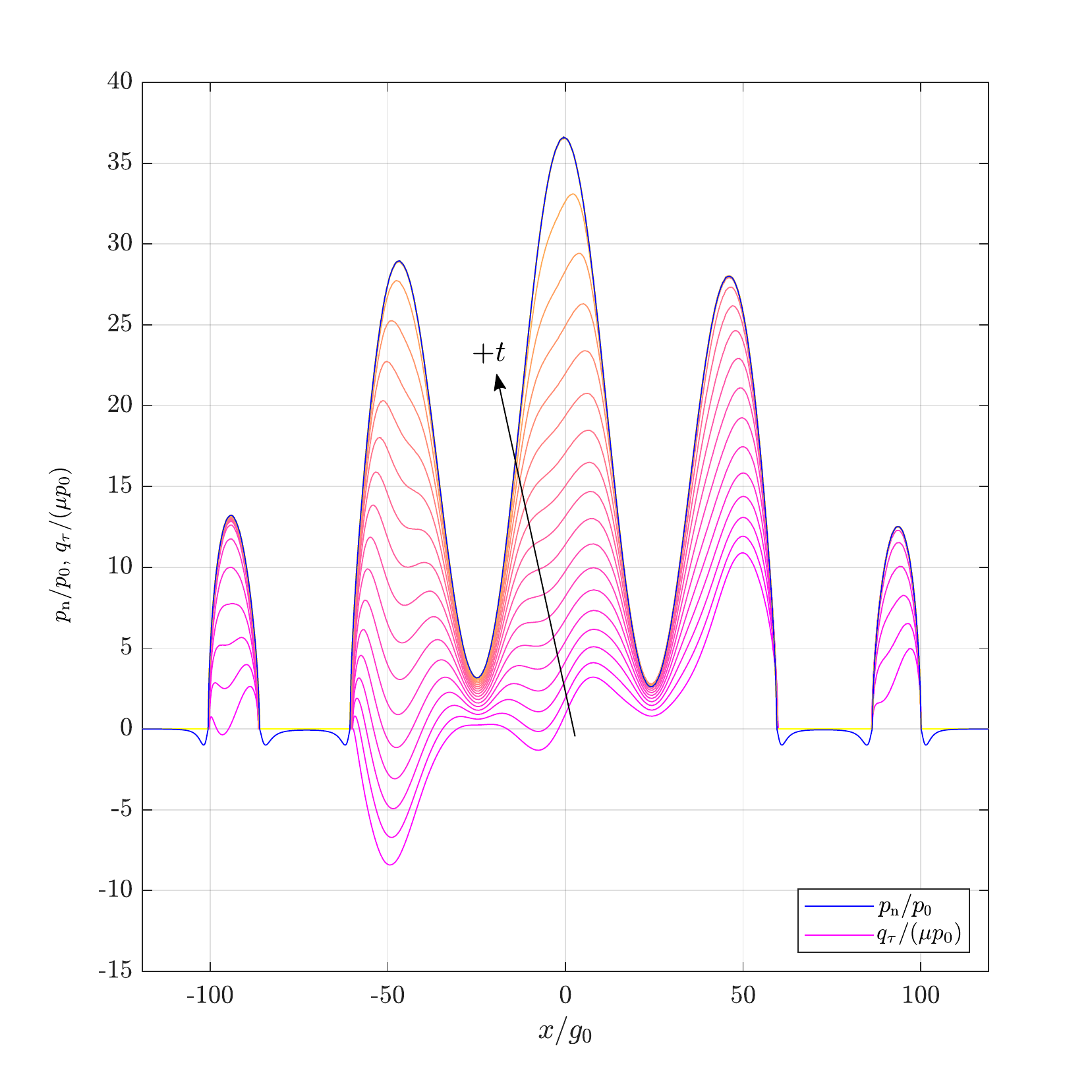}}
\caption{Example with friction, adhesion, and wavy profile.}
\label{fig:adh_fr}
\end{figure}

The simulation is set up under displacement control and solved in two different stages of equal length, each of them comprehensive of $25$ pseudo time steps spanning from $t=0$ to a unitary $t=\tf$. In the first phase, the profile is brought into contact by increasing a vertical far-field imposed displacement. The related solution is depicted in Fig.~\ref{fig:pzwma} in terms of normal contact tractions $\tn$, cyan ($t=0$) to blue ($t=\tf/2$) curves; and tangential tractions $q_\tau$, yellow ($t=0$) to magenta ($t=\tf/2$) curves. Both sets are normalized with respect to the highest value of the adhesive pressure, considered to be negative as opposed to positive contact tractions, as customary.

Each curve is related to a single pseudo-time step of the simulation and, as the indentation process advances, the three central asperities merge together forming a single cluster, while adjacent contact zones not yet connected are separated by depressed regions with a normal gap greater than $\gp$ but still displaying an appreciable effect of adhesive forces. 
In this phase, no tangential loading is applied, so that the related distribution of tangential forces is anti-symmetric and self equilibrated. 

In Fig.~\ref{fig:pzwmb}, results are shown when the horizontal motion is applied to the indenter, after fixing the vertical imposed displacement. Excluding slight variations due to normal-tangential coupling, normal tractions are now constant (solid blue line in the same figure) and the transition from a stick/slip regime to a full slip condition takes place, as can be seen from the final perfect overlapping between normal and tangential tractions scaled by $\mu$. \textcolor{black}{The evolution of tangential tractions can be traced together with the transition from magenta curves ($t=\tf/2$) to yellow curves ($t=\tf$).}

\subsection{3D simulations}
The approach formulated in Sec.~\ref{sec:derivation} is herein applied to 3D contact. The framework is first validated against the classic Hertz problem. Then, two bigger-scale applications involving complex surfaces under generic loading conditions are presented: $(i)$ frictionless normal contact of a\textcolor{black}{n} RMD rough surface; $(ii)$ contact between a rigid indenter characterized by a \emph{Weierstrass-Mandelbrot} self-affine surface, considering \textcolor{black}{the presence of friction at the interface and loading in the form of an} oblique far-field displacement. 

\subsubsection{Hertzian contact problem}
The Hertzian contact problem is used as a benchmark for the proposed 3D implementation. In the classic formulation of the problem, a paraboloid is employed as a first order approximation of a rigid spherical surface with radius $R=100\,\si{\milli\metre}$, which comes into contact with a deformable, linear elastic half-space. The problem is radially symmetric, and the solution is given in terms of contact radius $a$ and ellipsoidal normal contact tractions distributions $p(\mathbf{x})$, with $P$ being their resultant. Given a vertical imposed displacement $\Dn$, the aforementioned quantities read:
\begin{align}
a &= \sqrt{R\Delta_\mathrm{n}}, & P &= \frac{4}{3}\frac{E}{1-\nu^2}\sqrt{R\Delta_\mathrm{n}^3},
\end{align}
with $E$ and $\nu$ corresponding to Young's elastic modulus and Poisson's ratio of the half-space, respectively. The comparison is carried on under the application of a monotonically increasing vertical displacement, starting from zero up to a value of $\Delta_0 = 5\times 10^{-5} R$ with constant time steps. Finally, the values chosen for the bulk's characterization are $E = 1.0\,\si{\mega\pascal}$ and $\nu = 0.0$. Numerical simulations are performed, assuming both frictionless and frictional interfaces to highlight the differences that arise due to coupling and affect normal response also in absence of a direct tangential load.

\begin{figure}[b!]
\centering
\subfloat[][Bulk and interface discretization.\label{fig:hertz3Da}]
{\includegraphics[width=.45\textwidth]{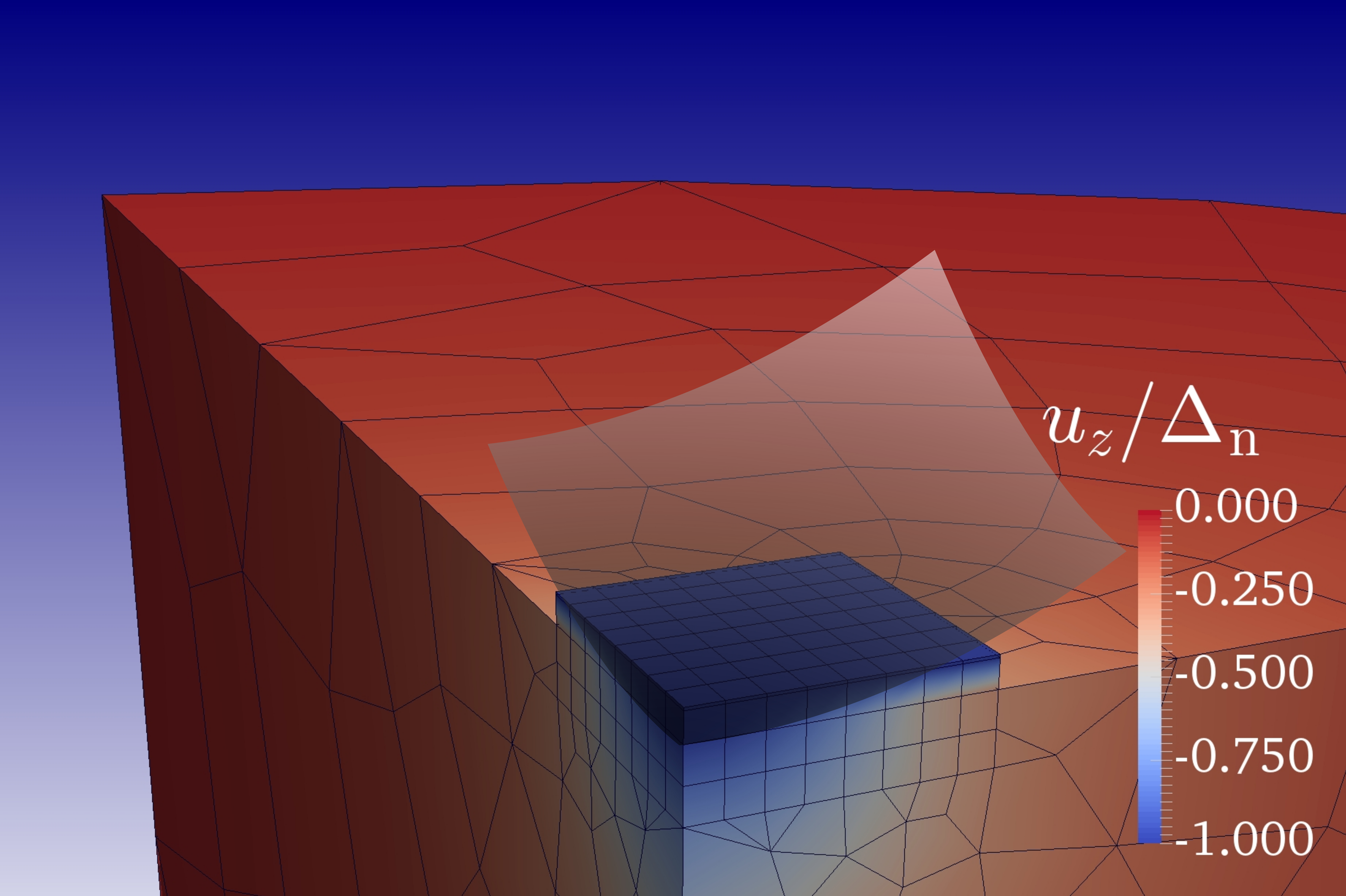}}\hspace{10mm}
\subfloat[][Contact patch magnification.\label{fig:hertz3Db}]
{\includegraphics[width=.45\textwidth]{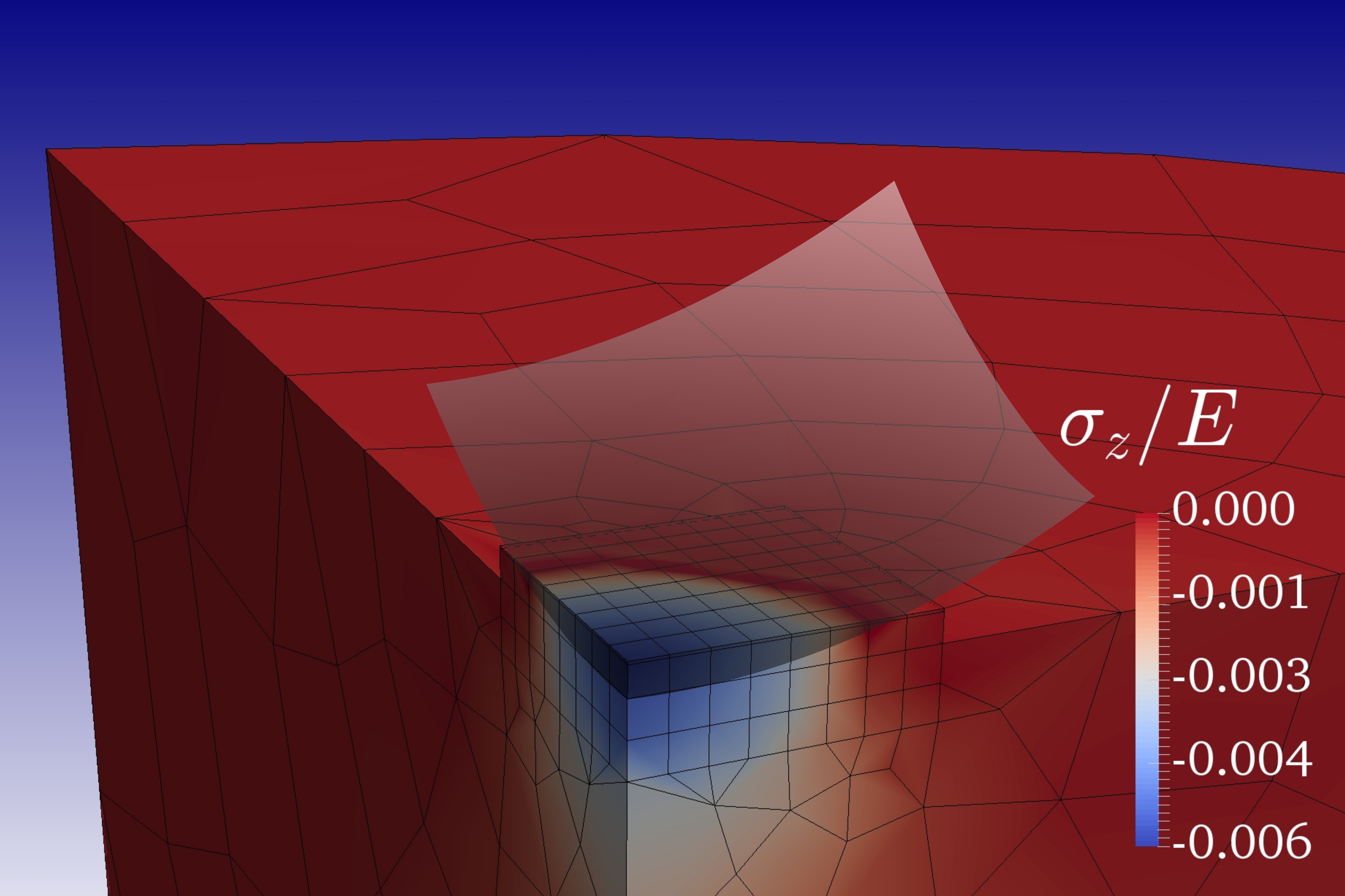}}
\caption{Problem set up characterized by deformable bulk and square contact patch of interface finite element. In the current case, a paraboloid surface is embedded in the contact elements, whose shape can be appreciated in transparency. BCs in the form of an imposed downward displacement are applied on top of the interface elements layer, and the resulting traction field is transmitted to the bulk. Contour plots show resulting vertical displacements $u_z$ (a) and resulting Cauchy stress $\sigma_z$ (b).}
\label{fig:hertz3D}
\end{figure}

Given the problem symmetry, only a quadrant of the half-space has been actually modeled and discretized with a quarter of cylinder, with rigid constraints in correspondence of the lower base and the round lateral surface and constraints in tangential direction on the two flat lateral surfaces. A layer of interface finite elements containing the shape of the indenting parabolic surface is located in correspondence to the top surface's center. Cylinder's radius and height have been increased until their influence on the simulation results vanished, thus guaranteeing the equivalence of the FEM bulk model response with the one expected for the half-space contact problem. A mesh convergence study has been performed regarding the discretization of the contact zone. Three different mesh resolutions have been employed, using square regular grids of $8\times8$, $16\times16$, $32\times32$ interface elements, respectively and a lateral size $L =1.0\times 10^{-2}R$. The problem setup can be appreciated in Fig.~\ref{fig:hertz3D} for the $8\times8$ resolution, while the results provided in the following paragraphs are all related to the fine resolution employed. Finally, a penalty parameter $\en = 1\times10^{8}E/R$ has been employed.

\begin{figure}[b!]
\centering
\subfloat[][Comparison with Hertz solution in terms of resulting force.\label{fig:Pz}]
{\includegraphics[width=0.3\textwidth]{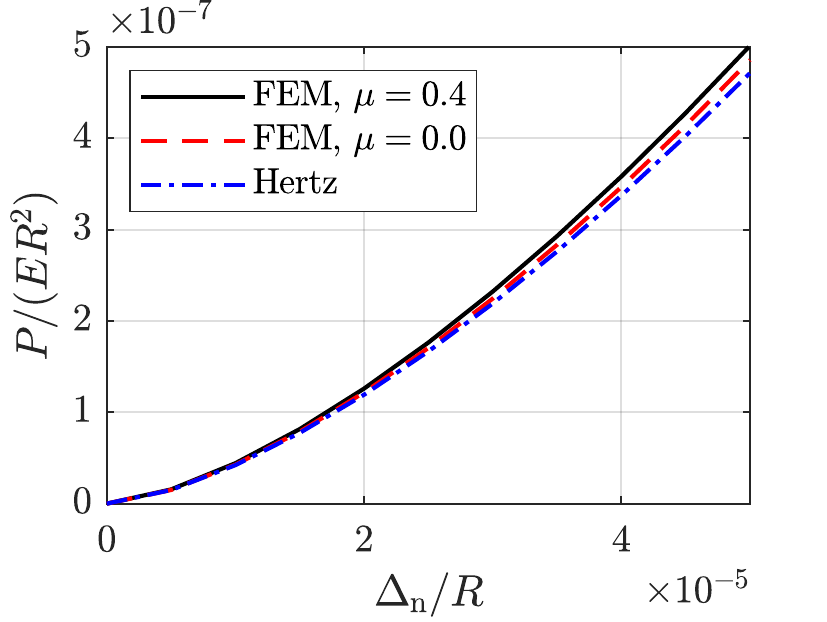}}
\subfloat[][Normal tractions field.\label{fig:tracpz_bench}]
{\includegraphics[width=0.3\textwidth]{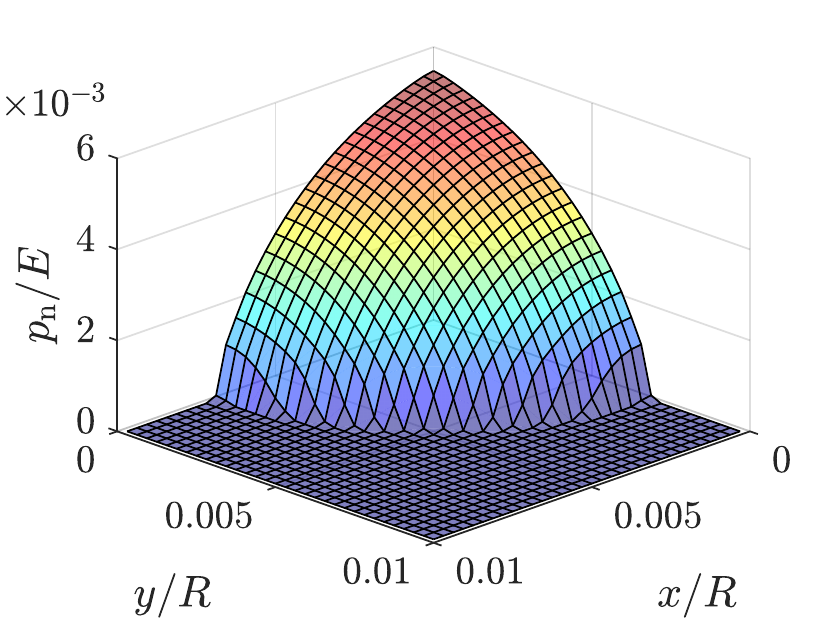}}
\subfloat[][Contour plot of $\tn$ with highlighted contact radius coming from the analytical solution.\label{fig:cradius}]
{\includegraphics[width=0.3\textwidth]{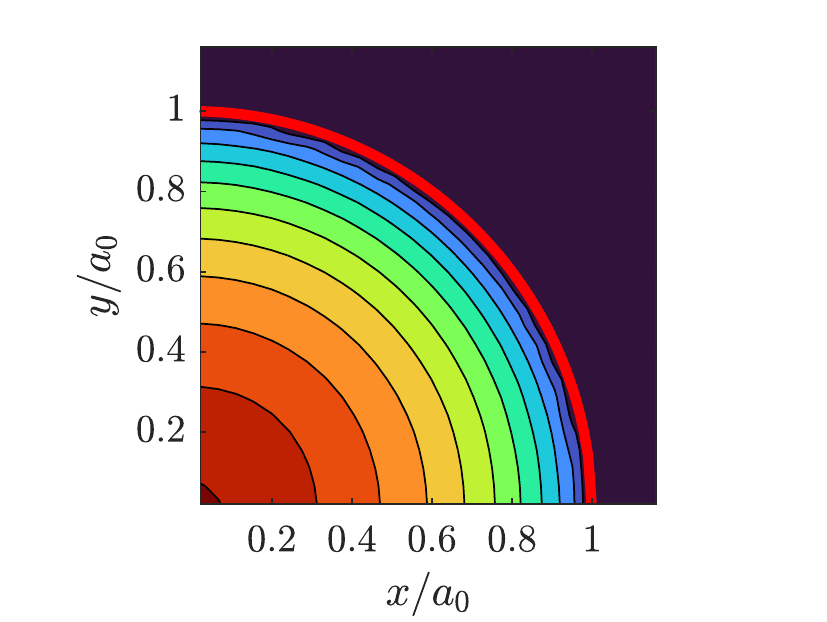}}
\caption{Comparison between numerical and analytical solutions for the Hertz problem.}
\label{fig:hertz}
\end{figure}

The solution in terms of contact reaction force, against the analytic reference solution, can be observed in Fig.~\ref{fig:Pz}, together with the surface plot of the correspondent normal tractions, Fig.~\ref{fig:tracpz_bench}. Both the frictionless ($\mu = 0.0$) and frictional ($\mu = 0.4$) numeric solutions show a stiffer behavior compared with the exact one. As expected, the frictional case is the stiffest since the application of the vertical load cause in-plane horizontal displacements, which are counteracted by the presence of friction. The highest coupling effect can be appreciated for $\nu=0.0$, while as Poisson's ratio tends to $0.5$, uncoupling conditions are met, and the effect is supposed to vanish. The differences in percentage between the case for $\mu = 0.4$ and $\mu = 0.0$ are in line with the theory. The interested reader is addressed to~\cite[Ch.\ 7, pp.\ 129--130]{barber:2018} for a comparison between the presented application and the corresponding coupled axis-symmetric problem without slip, which represents the scenario opposite to the absence of friction. A small but still appreciable difference still holds between the frictionless case and the reference solution. Even if the results of the validation test can be considered fully satisfactory since they have been obtained with a rather coarse mesh, the use of a different and more accurate contact strategy appears more appealing for the future systematic use of the method, for which the exploitation of the penalty based strategy is not a strict prerequisite.

In Fig.~\ref{fig:tracpz_bench} it can be seen how the numerical simulation reproduces the characteristic ellipsoidal shaped distribution. The stiffening effect due to geometrical coupling can be quantitatively appreciated by comparing the ratio between the maximum value predicted by the analytical model and the one obtained by the simulation, $p_0/p_\mathrm{max}=0.8105$. The solution in terms of contact radius is also checked. For the chosen interface discretization, a relative error of $1.612\%$ concerning the last loading step is found. This result is shown in Fig.~\ref{fig:cradius}, where the exact value of the contact radius, thick solid red line, is superposed to the normal contact tractions' contour plot.   

\begin{figure}[b!]
\centering
\subfloat[][$q_{\tau1}$ contact tractions distribution.\label{fig:tracqx}]
{\includegraphics[width=0.5\textwidth]{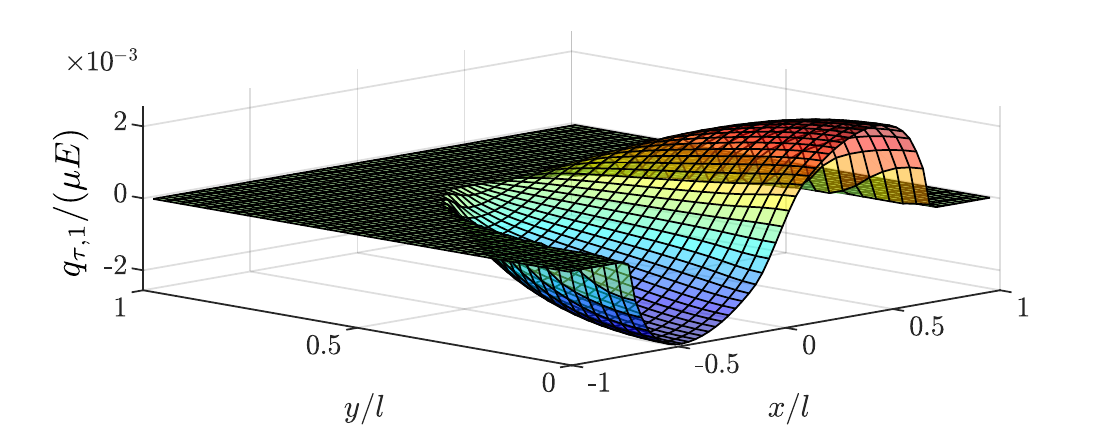}}
\subfloat[][Resulting $\norm{\mathbf{q}_\tau}$ distribution.\label{fig:tracq}]
{\includegraphics[width=0.5\textwidth]{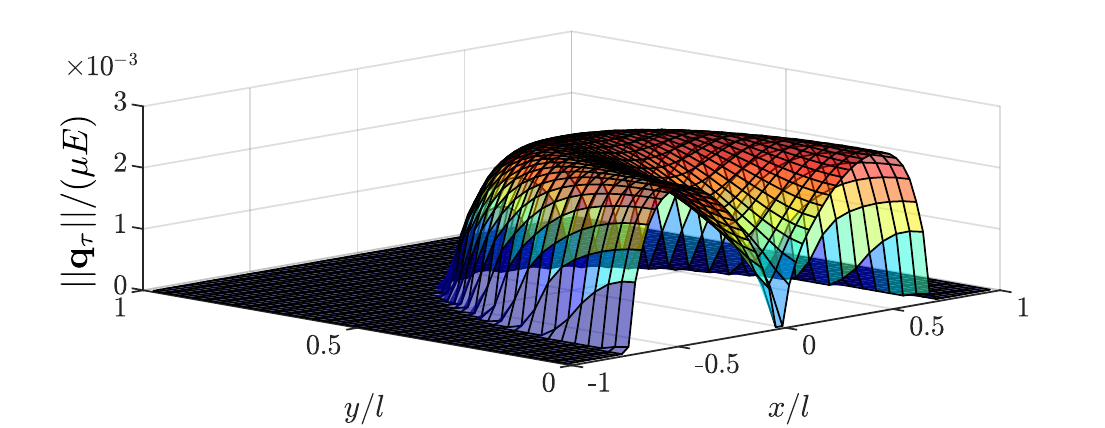}}
\caption{Surface plot of tangential tractions for the Hertz problem with friction.}
\label{fig:trac_tang}
\end{figure}

In Fig.~\ref{fig:trac_tang}, the tangential contact tractions are shown. Fig.~\ref{fig:tracqx} 
presents the tangent vector's projection over the first coordinate directions. Since only normal loading is involved, and the profile is symmetric, they represent self equilibrated distributions, symmetrical to $y=0$. The magnitude of the tangential tractions $\norm{\mathbf{q}_\tau}=\sqrt{q_{\tau,1}^2+q_{\tau,2}^2}$ is represented in Fig.~\ref{fig:tracq}. Again, because of loading conditions, the distribution is characterized by polar symmetry, with a null value only in correspondence to the origin. This point is the only one that does not experience in-plane tractions. The remaining domain is split in the radial direction into two annular regions, an inner one for which $\norm{\mathbf{q}_\tau}< \mu \tn$ which is therefore in a state of stick, and an outer one which radially slips under the action of the punch load. The stick/slip region can be determined by evaluating the ratio between the radius of stick and the contact radius, with the result $r_a/r_b = 0.9130$. This implies that roughly $15\%$ of the contact area is in a partial slip state, even for this rather high coefficient of friction and no application of tangential load. This fact might be of relevance in cases where micro-slip related phenomena are considered, for example, in the study of fretting wear and fretting fatigue~\cite{hills:1994,nowell:2006}.

\subsubsection{Contact of rough surfaces}
In this section, two different kinds of quasi-fractal rough surfaces are going to be tested, first a rough surface generated using an RMD algorithm, then a wavy Weierstrass-Mandelbrot (WD) quasi fractal surface. Two different methodologies have been used for the assignment of the correct elevation field to each elements' Gau\ss~points. The surface employed in the contact simulation \textcolor{black}{has been hardcoded inside the finite element routine in the case of the WD surface since it can be analytically defined. In the first case, it has been stored in an external file as a three columns matrix of $[x,y,z]$ values and prompted as a look-up table, this solution is necessary in the case the surface to be used directly comes from topographic measurements, such as those obtained from a confocal profilometer, or if, in general, it lacks an analytical description as is the case for the RMD surface.}

In analogy with Sec.~\ref{sec:RMD}, the first case is considered particularly interesting, since a contact problem involving this type of surface can be particularly challenging when using standard contact search algorithms, given the scatter in the heights distribution and the total lack of smoothness.


Each of the two simulations is performed over the same mesh, \textcolor{black}{which is structured over three different layers stacked on the top of each other. The bottom layer models the bulk, the middle layer is composed of interface finite elements where the indenter's geometry is sampled and finally, a top layer where Dirichlet BCs are applied, according to the scheme depicted in Fig.~\ref{fig:bcsb}}. Standard trilinear \emph{hex} elements have been employed for modeling the first and the last layer.

\begin{figure}[t!]
\centering
\includegraphics[width=0.5\textwidth]{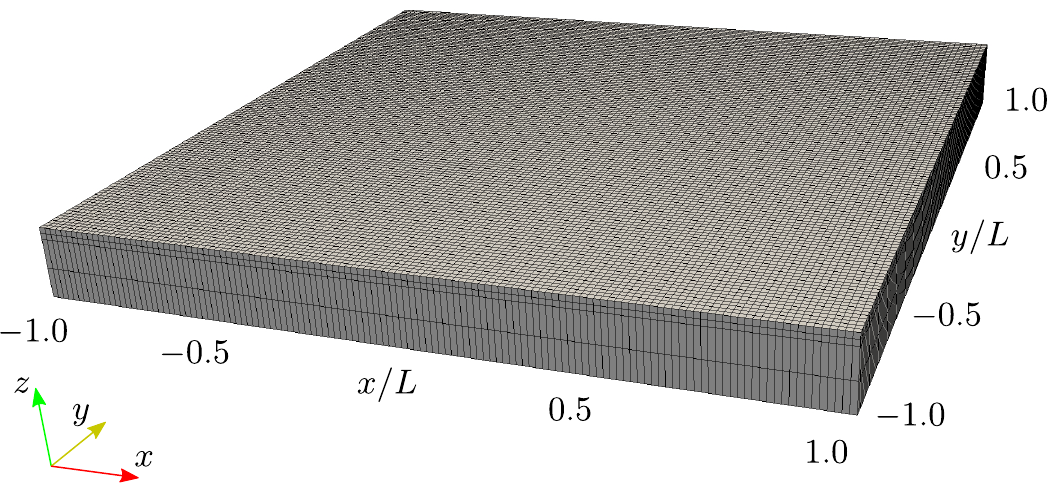}
\caption{FEM mesh, interface discretized with $128\times128$ interface finite elements.}
\label{fig:fine_mesh}
\end{figure}

The indenter\textcolor{black}{s are sampled} in an array composed of $128\times128$ square interface finite elements. Bulk elements have a height-to-width ratio of $5$, which, given the square nominal contact area of side $2L$ with $L=1\,\si{\milli\metre}$ and the number of elements employed, gives an overall depth of $h_\mathrm{b} = 0.1563L$. Fig.~\ref{fig:fine_mesh} shows an overview of the mesh  employed. The problem setup is completed by its mechanical characterization. The bulk is considered to be linear elastic, with Young's modulus $E = 1.0\,\si{\megapascal}$ and a Poisson's ratio $\nu = 0.0$, while the material predisposed for the application of the BCs is considered three orders of magnitude stiffer than the bulk. The normal penalty parameter has been taken to be $\en = 1\times10^{3}E/h_\mathrm{b}$. Finally, no restraints are considered on the free lateral surfaces of the elastic bulk.

\begin{figure}[t!]
\centering
\subfloat[][RMD surface.\label{fig:RMDsurfa}]
{\includegraphics[width=0.5\textwidth,valign=c]{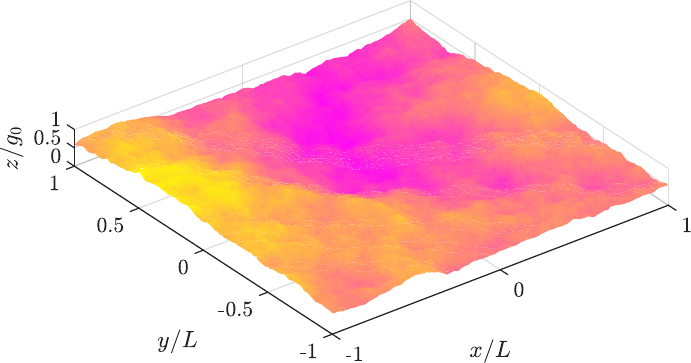}}
\subfloat[][Contour plot of normal tractions $p(\mathbf{x})$.\label{fig:RMDpz}]
{\includegraphics[width=0.5\textwidth,valign=c]{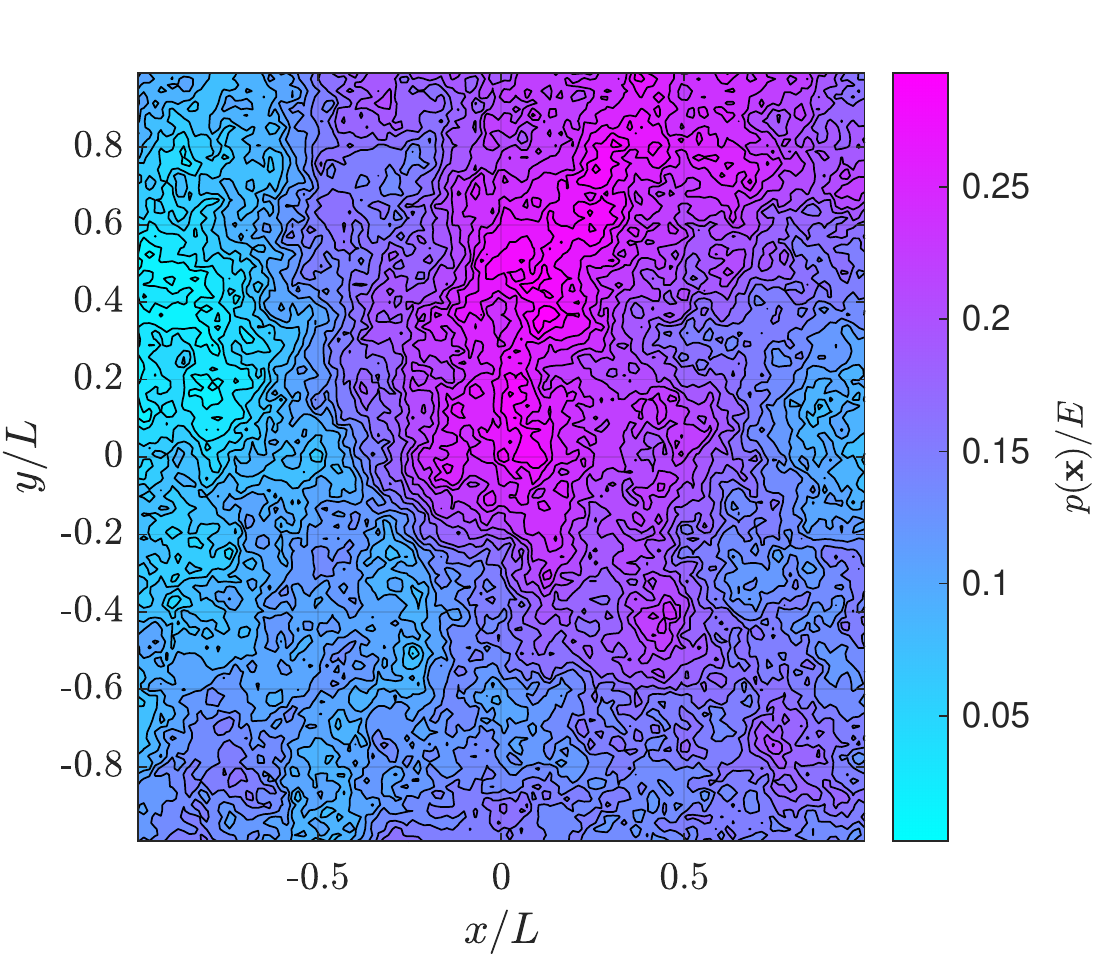}}
\caption{Surface employed in 3D simulations and resultant traction field.}
\label{fig:surf_WM_RMD}
\end{figure}

\paragraph{Results for RMD surface}
A self-affine rough surface obtained employing the same procedure of Sec.~\ref{sec:RMD} is now used for testing the $3D$ implementation. The surface is generated with the RMD algorithm and a fixed random seed $r = 0.547$, a Hurst exponent $H = 0.75$ and a random function with Gaussian distribution and a starting standard deviation $\sigma_0 = 2.357$. The resulting elevation field is shown qualitatively in Fig.~\ref{fig:RMDsurfa}. 

\begin{figure}[h!]
\centering
\subfloat[][Far-field/Contact force.\label{fig:RMDload}]
{\includegraphics[width=0.4\textwidth]{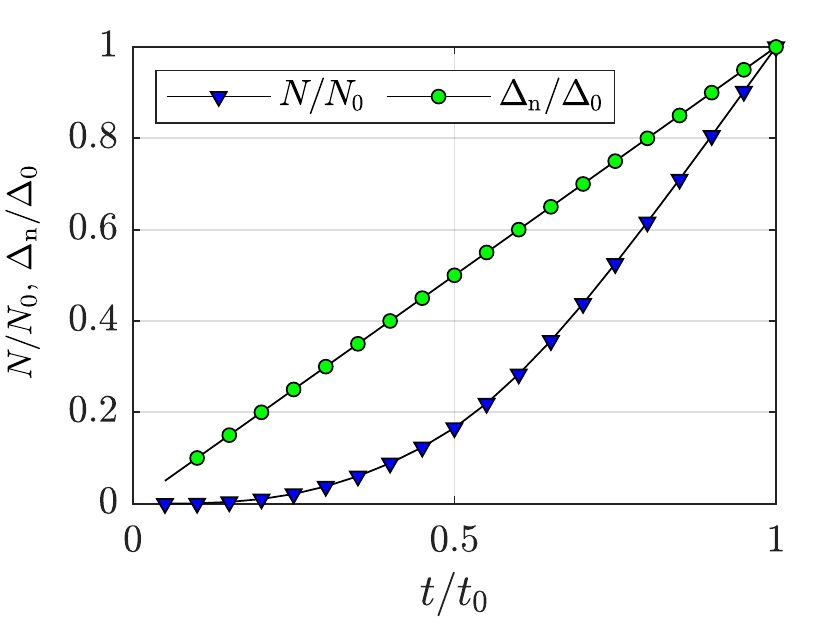}}\hspace{5mm}
\subfloat[][Contact area.\label{fig:RMDac}]
{\includegraphics[width=0.4\textwidth]{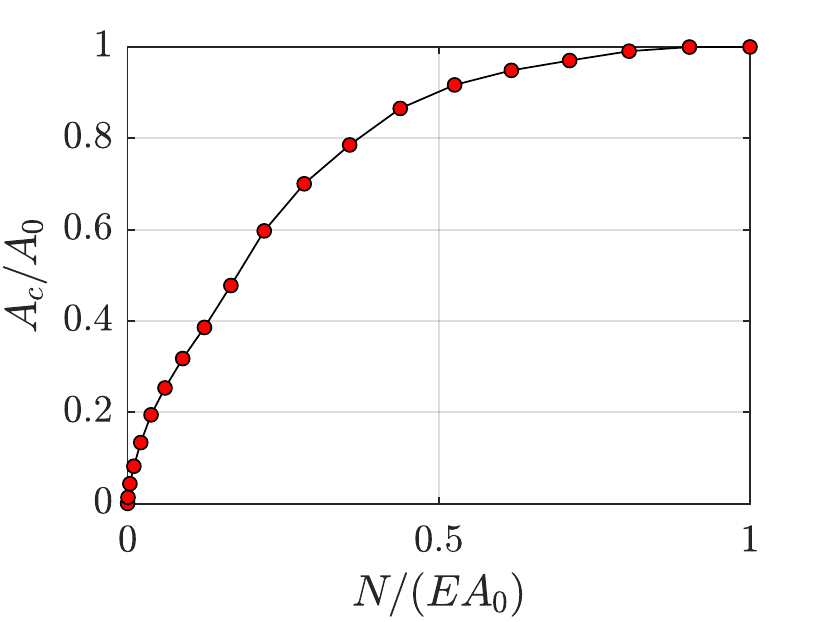}}
\caption{Solution of the indentation problem for an RMD fractal surface.}
\label{fig:RMD}
\end{figure}

\textcolor{black}{In this case, a frictionless normal indentation problem is solved. The load is applied as an imposed far field displacement $\Dn$ on the top layer of rigid elements, linearly varying from a null value up to a maximum of $\Dz = g_0 = 1.0\times10^{-2}L$, discretized employing $20$ pseudo time step. Again, $g_0$ represents the amplitude of the surface measured from the lowest valley to the highest summit.}
The load history is plotted in Fig.~\ref{fig:RMDload} in terms of imposed normal far field displacement $\Dn$, together with the resultant normal reaction force $P$, scaled by their maximum values $\Dz$ and $P_0 = 0.637 E/L^2$, with $\tf$ the final instant of the simulation. The maximum value of the imposed displacement has been chosen high enough to map the evolution of the actual contact area $\Ac$, from a single contacting asperity at $t=0$ to full contact at $t=\tf$, as can be seen in Fig.~\ref{fig:RMDac} where this quantity scaled by the nominal contact area $A_0 = 4L^2$ is plotted. Finally, the contour plot of the full normal tractions field is reported in Fig.~\ref{fig:RMDpz}, with a peak value $p_0=0.308E$.

\begin{figure}[t!]
\centering
\subfloat[][WM surface.\label{fig:WMsurf}]
{\includegraphics[width=0.5\textwidth,valign=c]{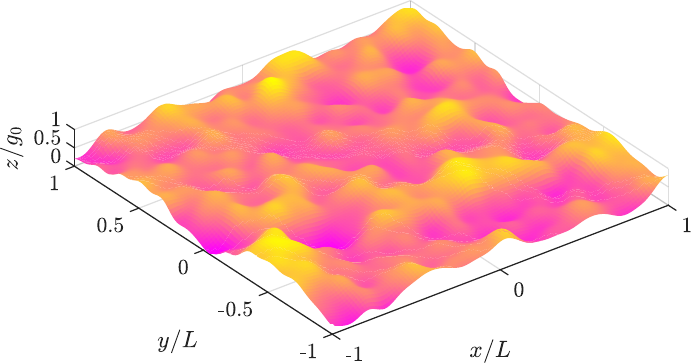}}
\subfloat[][Contour plot of normal tractions $p(\mathbf{x})$.\label{fig:WMpz}]
{\includegraphics[width=0.5\textwidth,valign=c]{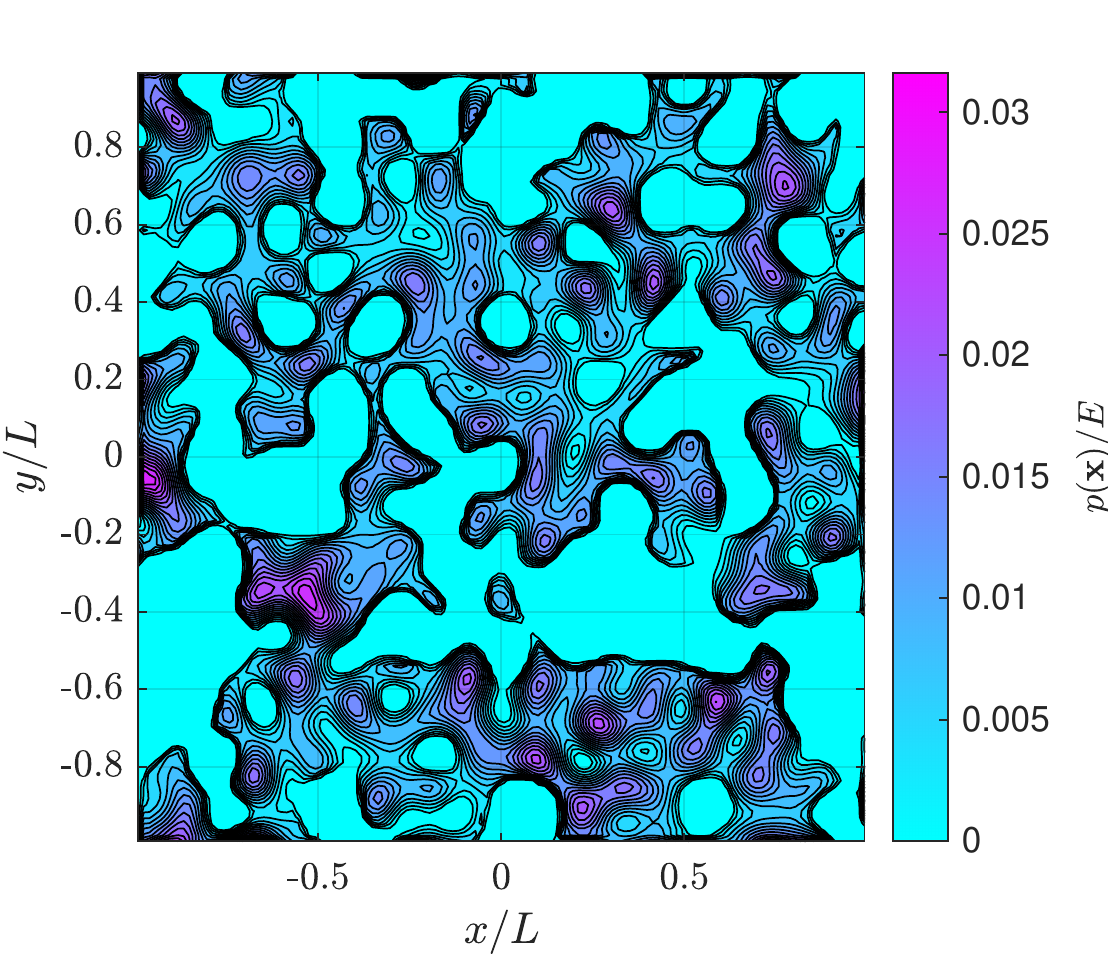}}
\caption{Surface employed in 3D simulations and resultant traction field.}
\end{figure}

\paragraph{WM with friction}
\textcolor{black}{The second full scale simulation is performed considering the presence of friction at the interface, with a coefficient of friction $\mu=0.2$. The indenter's surface is a quasi-fractal} Weierstrass-Mandelbrot surface~\cite[Ch.\ 16, pp.\ 356]{fractals:1988,mandelbrot:1977,barber:2018} defined by the function:
\begin{equation}
z(x,y) = A \sum_{n=1}^N \sum_{m=1}^M \gamma^{(D-3)(n-1)}\Bigl[\cos{\phi_{m,n}}
-\cos{\frac{2\pi\gamma^{n-1}}{\lambda_0}}\Bigl(x\cos{\frac{\pi m}{M}}+y\sin{\frac{\pi m}{M}}+\phi_{m,n}\Bigl)\Bigl],
\label{eq:WM}
\end{equation}
and characterized by the parameters in Tab.~\ref{tab:coeffWM} and shown in Fig.~\ref{fig:WMsurf}. The matrix $\Phi$ collects the random phase angles employed for the surface generation process.

\begin{table}[b!]
\centering
\caption{Weierstrass-Mandelbrot surface coefficients.}
{\footnotesize
\begin{tabular}{ccccccc}
\toprule
$z_0$ & $\lambda_0$ & $G$ & $D$ & $\gamma$ & $N$ & $M$\\ 
\midrule
$[\si{\metre}]$ & $[\si{\metre}]$ & $[-]$ & $[-]$ & $[-]$ & $[-]$ & $[-]$\\
\midrule
$1.00\times10^{-3}$ & $1.00\times10^0$ & $3.00\times10^0$ & $2.25\times10^0$ & $1.30\times10^0$ & $8$ & $10$\\
\bottomrule
\label{tab:coeffWM}
\end{tabular}}
\end{table}

\begin{figure}[t!]
\centering
\includegraphics[width=0.8\textwidth]{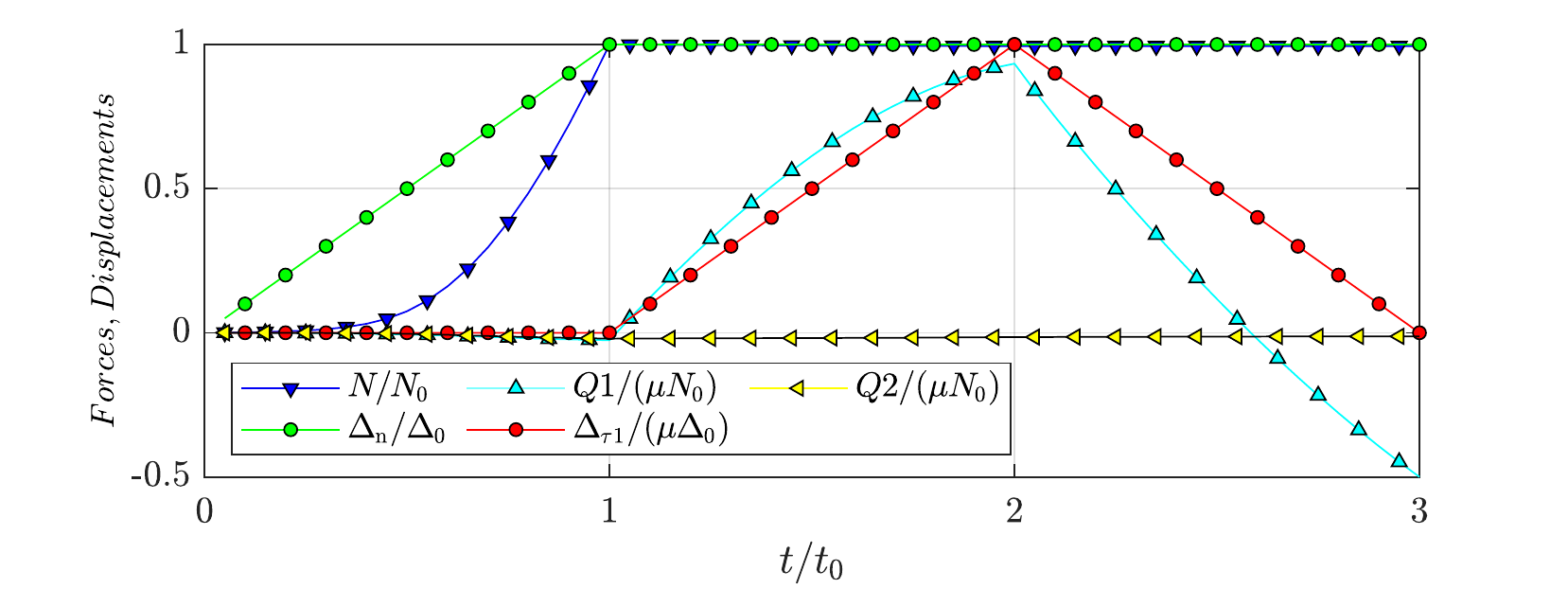}
\caption{Far field displacement and resultant load vs. time.}
\label{fig:farf}
\end{figure}

\textcolor{black}{The load is still applied as a far field displacement on the top layer of the mesh, this time considering also horizontal imposed motion. The overall loading phase is considered quasi-static and discretized in $60$ pseudo time steps, ranging from zero to $3t_0$. The overall loading process is divided into three different stages. In the first, ranging from zero to $t_0$, a pure vertical displacement is applied from a null value up to $\Delta_0 = 3.0\times10^{-1}g_0$. The normal displacement is then held constant, while the indenter is shifted along $x$ direction with constant positive velocity, reaching a maximum value $\Delta_{\tau,0}=\mu\Delta_0$ at $2t_0$. Finally, in the third phase, the indenter is linearly shifted back to its original position, reached at $3t_0$. Figure~\ref{fig:farf} shows the applied far-field displacement history, together with the resultant interface overall reactions, evaluated as the integral of the interface normal and tangential tractions.}

The ratio between the normal indentation and the elastic layer thickness is $\Delta_0/h_\mathrm{b}=1.92\,\%$, in line with the assumption of elastic deformation of the bulk. Still, the surface characteristics have been tailored to obtain a high final actual contact area to have the possibility of investigating the contact response from high to low mean-plane separations.

Considering the WM related simulations, the outcome in terms of forces response is also shown in Fig.~\ref{fig:farf}. The vertical reaction force $P$ follows a characteristic power-law behavior as long as the load is incremented, then remains constant. During the first stage, parasitic reaction forces $Q_1$ and $Q_2$ arise due to the simulation's displacement controlled nature and the lack of symmetry of the indenting profile. During the second stage, $Q_1$ increases, and a condition of full slip is almost reached, with the maximum value obtained at $2t_0$ approximately equal to $0.85\mu P$. Over this point, the displacement is reversed, and the indenter is taken back to its original position. We observe a residual horizontal negative force, a function of the system hysteresis that can be directly linked to the frictional energy dissipation.

The contour plot of the normal tractions $p(\mathbf{x})$ at $t_0$ is shown in Fig.~\ref{fig:WMpz}. It can be seen that for the selected level of indentation, the contact area ratio $\Ac/A_0\simeq 45\%$ is reached. A clear distinction holds between the contact islands and the domain that does not experience contact, characterized by homogeneous cyan color.

\subsubsection{Computational performances}

\begin{figure}[b!]
\centering
\subfloat[][CPU time.\label{fig:cpu}]
{\includegraphics[width=0.5\textwidth]{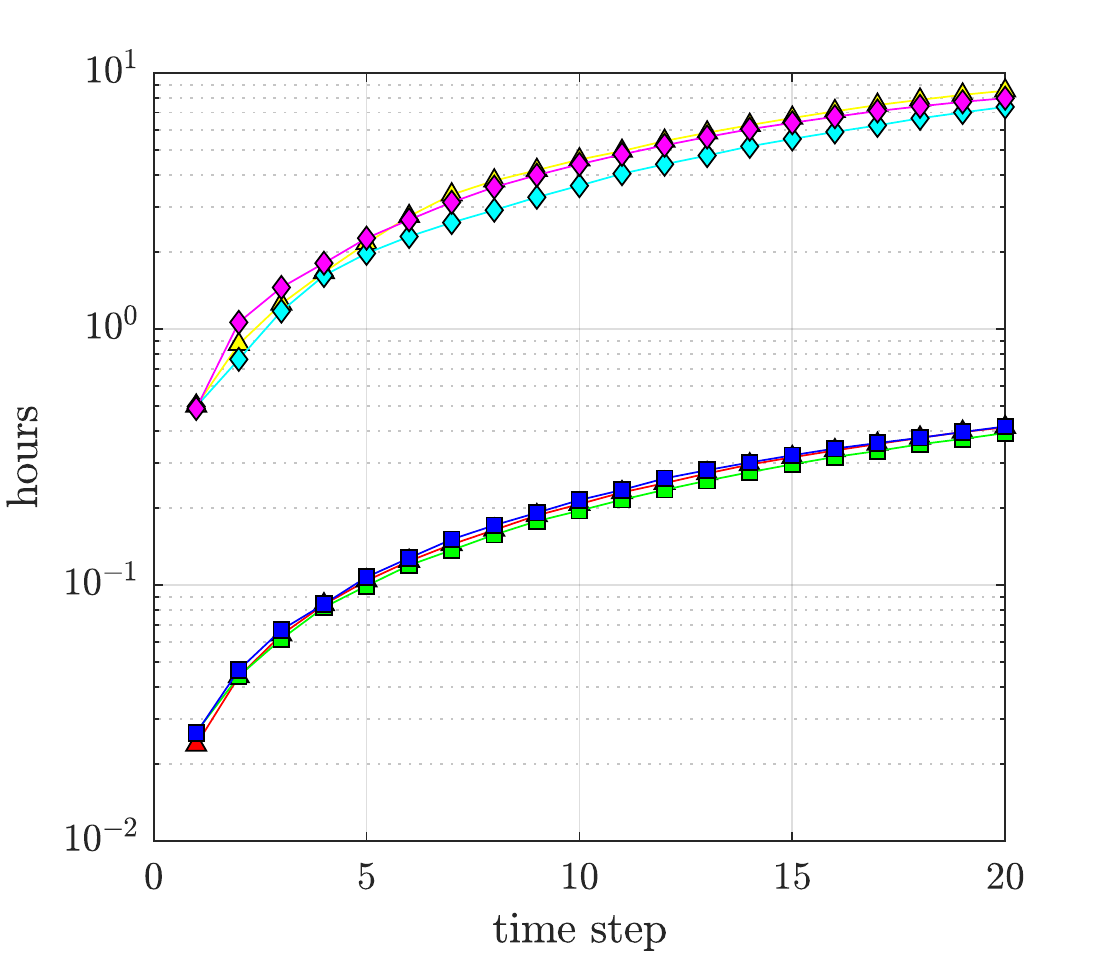}}
\subfloat[][Number of Newton-Raphson iterations necessary for reaching convergence.\label{fig:nr}]
{\includegraphics[width=0.5\textwidth]{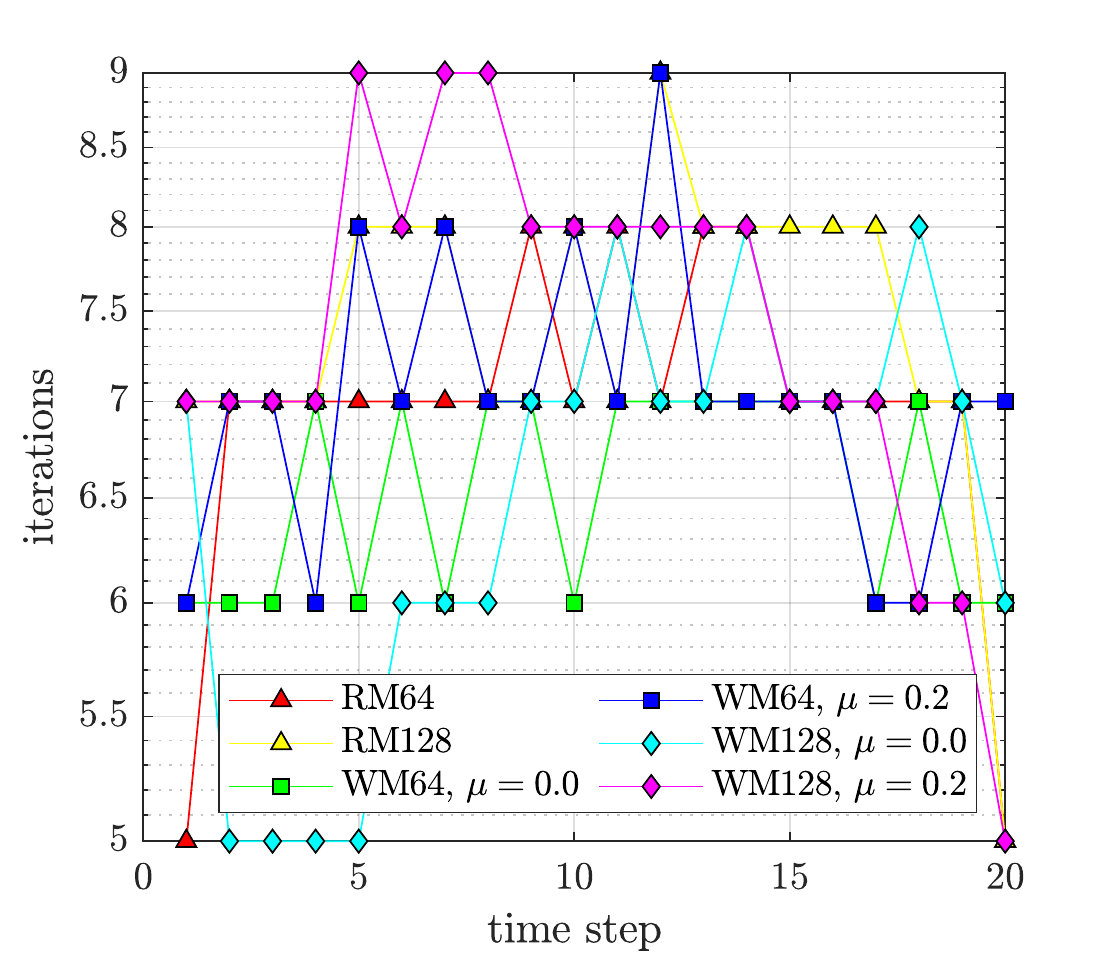}}
\caption{Comparison of the solver performances between the two full scale examples addressed, together with an analogous problem characterized by a lower number of degrees of freedom.}
\label{fig:cpunr}
\end{figure}

\textcolor{black}{Results for both the RMD and the WM surfaces are compared in terms of computational time required and convergence properties at the end of this section. The performance of the proposed method is compared for both the RMD and the WM surfaces, along the first load branch, i.e. from zero to the $20^{th}$ time step. Each simulation ran sequentially on the single node of an Intel Xeon E5$\cdot$4620 processor with $256\,\si{\giga\byte}$ of ram. In the solution process, a full Newton-Raphson solution scheme together with a direct solver based on Gau{\ss}ian elimination for the inversion of the global tangent stiffness matrix has been employed. For the simulations involving friction, an implicit backward Euler time stepping scheme has been employed, while dynamical forces have not been taken into account.}

Figure~\ref{fig:cpu} shows the time employed by a complete run of all the simulations performed. For comparison purposes, results related to solutions obtained employing a lower number of degrees of freedom (surfaces modeled on $64\times64$ elements grids) are plotted as well in the same figure. As expected, the most critical factor is the number of degrees of freedom that characterizes the different examples. For what concerns finer scale problems, all the simulations with an equivalent number of degrees of freedom have almost identical CPU times, regardless of the surfaces' smoothness. In contrast to the WM surface, the RMD surface is made of a scattered elevation field, which would result in very challenging scenarios for standard contact search algorithms. In order to investigate how the presence of friction affects the performance of the code, the same problem with the WM surface is solved also setting $\mu=0.0$. Comparing the results, a slight difference is encountered, but the effect is noticeable for the finest resolution only, with an increase of about $12\%$ concerning the overall computational time, and convergence properties as well are not significantly affected. In the conclusion of the section, Fig.~\ref{fig:nr} reports, for each time step of each simulation, the total number of iterations of the Newton-Raphson algorithm employed to solve the global non-linear system of equations that governs the problem. Again, no significant discrepancy is encountered despite the remarkable differences in terms of smoothness characteristics. Furthermore, in the case of the WM surface, even friction does not significantly alter the convergence properties, requiring at most two additional iterations for reaching convergence.



\section{Conclusion and future perspectives}\label{sec:conclusion}
\textcolor{black}{In this paper, an extension to the MPJR interface finite element is presented for the analysis of rough 2D and 3D contact problems. Good accordance has been found comparing the proposed implementation with solutions obtained from standard numerical frameworks for the solution of the frictionless normal contact problem of a rough RMD indenting profile. The setup proved to be valid also for the analysis of contact problems with wavy interfaces in presence of friction and adhesion.}
The proposed formulation provides a way to overcome some of the major difficulties related to the solution of contact problems with roughness in the state-of-the-art BEM and FEM formulations.
With respect to classical BEM solvers, the proposed method allows considering any nonlinear constitutive relation for the bulk and for the interface. Moreover, it  allows simulating finite size geometries and it is naturally prone to be extended for multi-field simulations (phase field for fracture mechanics in the bulk, thermo-elasticity, chemical reactions coupled with mechanics, etc.). 

As far as standard FEM procedures are concerned, the methodology simplifies the discretization of the interface, which does not need to be explicitly included in the model geometry. This allows including any point-wise height field or any analytical shape of 2D profiles or 3D surfaces as a straightforward correction to the normal gap. In the case of simulations based on experimentally acquired profile/surface data (with AFM, confocal profilometer, or any other technique), the height field can be efficiently stored into a history variable which is then compiled with the code and read by the FE software only once at the initialization stage of the problem. This avoids repeated read-write operations from external files.

Using the proposed formulation, the contacting interface is treated as nominally flat and roughness is embedded at the interface level node-wise. Therefore, the method requires \textcolor{black}{an interface discretization that is consistent with the number of data points required for accurately sampling the indenter's boundary, together with their spacing}. This allows for an exact reproduction of the contacting geometry by using low-order interpolation schemes, without compromising the convergence that can be a problem for contact search algorithms with not well-defined normal vectors.

Future perspectives comprehend the consistent application of the method to model full-scale 3D contact problems under finite strain assumptions for the study of phenomena where surface roughness plays a key role as in wear problems, fracture-induced by indentation, fracture induced by repeated application of contact loads, tire-asphalt interaction, nanoscale tribological tests based on AFM data, multi-field tribological problems. 

Finally, the authors are grateful to Jim Barber for the frequent scientific discussion they had with him over their entire careers. This has inspired (and we hope will continue to inspire) many research ideas on contact mechanics between rough surfaces that would have not been possible to pursue without its input. 


\section*{Acknowledgements}
JB and MP would like to thank the Italian Ministry of Education, University and Research (MIUR) for the support to the Research Project of Relevant National Interest (PRIN 2017) XFAST-SIMS: Extra-fast and accurate simulation of complex structural systems (Grant agreement n. 20173C478N).
DD would like to acknowledge the support received from the Engineering and Physical Science Research Council (EPSRC) via his Established Career Fellowship (EP/N025954/1).

\section*{Declaration of interests}
The authors declare that they have no known competing financial interests or personal relationships that could have appeared to influence the work reported in this paper.

\bibliography{ref}

\begin{thebibliography}{10}
\expandafter\ifx\csname url\endcsname\relax
  \def\url#1{\texttt{#1}}\fi
\expandafter\ifx\csname urlprefix\endcsname\relax\def\urlprefix{URL }\fi
\expandafter\ifx\csname href\endcsname\relax
  \def\href#1#2{#2} \def\path#1{#1}\fi

\bibitem{barber:1968}
J.~Barber, Thermal effects of friction and wear, Ph.D. thesis, University of
  Cambridge (1968).

\bibitem{ahn:2008a}
Y.~Ahn, J.~Barber,
  \href{https://linkinghub.elsevier.com/retrieve/pii/S0020740308001203}{Response
  of frictional receding contact problems to cyclic loading}, International
  Journal of Mechanical Sciences 50~(10-11) (2008) 1519--1525.
\newblock \href {https://doi.org/10.1016/j.ijmecsci.2008.08.003}
  {\path{doi:10.1016/j.ijmecsci.2008.08.003}}.
\newline\urlprefix\url{https://linkinghub.elsevier.com/retrieve/pii/S0020740308001203}

\bibitem{ahn:2008b}
Y.~Ahn, E.~Bertocchi, J.~Barber,
  \href{https://linkinghub.elsevier.com/retrieve/pii/S0022509608001488}{Shakedown
  of coupled two-dimensional discrete frictional systems}, Journal of the
  Mechanics and Physics of Solids 56~(12) (2008) 3433--3440.
\newblock \href {https://doi.org/10.1016/j.jmps.2008.09.003}
  {\path{doi:10.1016/j.jmps.2008.09.003}}.
\newline\urlprefix\url{https://linkinghub.elsevier.com/retrieve/pii/S0022509608001488}

\bibitem{barber:2011a}
J.~Barber, M.~Davies, D.~Hills,
  \href{https://linkinghub.elsevier.com/retrieve/pii/S0020768311001041}{Frictional
  elastic contact with periodic loading}, International Journal of Solids and
  Structures 48~(13) (2011) 2041--2047.
\newblock \href {https://doi.org/10.1016/j.ijsolstr.2011.03.008}
  {\path{doi:10.1016/j.ijsolstr.2011.03.008}}.
\newline\urlprefix\url{https://linkinghub.elsevier.com/retrieve/pii/S0020768311001041}

\bibitem{barber:1969}
J.~Barber, W.~Hawthorne, M.~Lighthill,
  \href{https://royalsocietypublishing.org/doi/abs/10.1098/rspa.1969.0165}{Thermoelastic
  instabilities in the sliding of conforming solids}, Proceedings of the Royal
  Society of London. A. Mathematical and Physical Sciences 312~(1510) (1969)
  381--394.
\newblock \href
  {http://arxiv.org/abs/https://royalsocietypublishing.org/doi/pdf/10.1098/rspa.1969.0165}
  {\path{arXiv:https://royalsocietypublishing.org/doi/pdf/10.1098/rspa.1969.0165}},
  \href {https://doi.org/10.1098/rspa.1969.0165}
  {\path{doi:10.1098/rspa.1969.0165}}.
\newline\urlprefix\url{https://royalsocietypublishing.org/doi/abs/10.1098/rspa.1969.0165}

\bibitem{barber:1971}
J.~Barber,
  \href{https://linkinghub.elsevier.com/retrieve/pii/0017931071901050}{The
  effect of thermal distortion on constriction resistance}, International
  Journal of Heat and Mass Transfer 14~(6) (1971) 751--766.
\newblock \href {https://doi.org/10.1016/0017-9310(71)90105-0}
  {\path{doi:10.1016/0017-9310(71)90105-0}}.
\newline\urlprefix\url{https://linkinghub.elsevier.com/retrieve/pii/0017931071901050}

\bibitem{barber:1976}
J.~Barber,
  \href{https://academic.oup.com/qjmam/article-lookup/doi/10.1093/qjmam/29.1.1}{{SOME}
  {THERMOELASTIC} {CONTACT} {PROBLEMS} {INVOLVING} {FRICTIONAL} {HEATING}}, The
  Quarterly Journal of Mechanics and Applied Mathematics 29~(1) (1976) 1--13.
\newblock \href {https://doi.org/10.1093/qjmam/29.1.1}
  {\path{doi:10.1093/qjmam/29.1.1}}.
\newline\urlprefix\url{https://academic.oup.com/qjmam/article-lookup/doi/10.1093/qjmam/29.1.1}

\bibitem{barber:2003}
J.~Barber,
  \href{https://royalsocietypublishing.org/doi/10.1098/rspa.2002.1038}{Bounds
  on the electrical resistance between contacting elastic rough bodies},
  Proceedings of the Royal Society of London. Series A: Mathematical, Physical
  and Engineering Sciences 459~(2029) (2003) 53--66.
\newblock \href {https://doi.org/10.1098/rspa.2002.1038}
  {\path{doi:10.1098/rspa.2002.1038}}.
\newline\urlprefix\url{https://royalsocietypublishing.org/doi/10.1098/rspa.2002.1038}

\bibitem{barber:2013a}
J.~Barber, \href{http://link.springer.com/10.1007/s11249-012-0094-6}{Multiscale
  {Surfaces} and {Amontons}’ {Law} of {Friction}}, Tribology Letters 49~(3)
  (2013) 539--543.
\newblock \href {https://doi.org/10.1007/s11249-012-0094-6}
  {\path{doi:10.1007/s11249-012-0094-6}}.
\newline\urlprefix\url{http://link.springer.com/10.1007/s11249-012-0094-6}

\bibitem{barber:2013b}
J.~Barber,
  \href{https://link.aps.org/doi/10.1103/PhysRevE.87.013203}{Incremental
  stiffness and electrical contact conductance in the contact of rough finite
  bodies}, Physical Review E 87~(1) (2013) 013203.
\newblock \href {https://doi.org/10.1103/PhysRevE.87.013203}
  {\path{doi:10.1103/PhysRevE.87.013203}}.
\newline\urlprefix\url{https://link.aps.org/doi/10.1103/PhysRevE.87.013203}

\bibitem{paggi:2011c}
M.~Paggi, J.~R. Barber, Contact conductance of rough surfaces composed of
  modified rmd patches, International Journal of Heat and Mass Transfer 54
  (2011) 4664--4672.
\newblock \href {https://doi.org/10.1016/j.ijheatmasstransfer.2011.06.011}
  {\path{doi:10.1016/j.ijheatmasstransfer.2011.06.011}}.

\bibitem{barber:2002}
J.~Barber, \href{https://books.google.de/books?id=qUht0jvZcIoC}{Elasticity},
  Online access with purchase: {Springer}, Springer Netherlands, 2002.
\newline\urlprefix\url{https://books.google.de/books?id=qUht0jvZcIoC}

\bibitem{barber:2011b}
J.~Barber, \href{https://books.google.de/books?id=qUht0jvZcIoC}{Intermediate
  Mechanics of Materials}, Springer Netherlands, 2011.
\newline\urlprefix\url{https://books.google.de/books?id=qUht0jvZcIoC}

\bibitem{barber:2018}
J.~Barber, \href{http://link.springer.com/10.1007/978-3-319-70939-0}{Contact
  {Mechanics}}, Vol. 250 of Solid {Mechanics} and {Its} {Applications},
  Springer International Publishing, Cham, 2018.
\newblock \href {https://doi.org/10.1007/978-3-319-70939-0}
  {\path{doi:10.1007/978-3-319-70939-0}}.
\newline\urlprefix\url{http://link.springer.com/10.1007/978-3-319-70939-0}

\bibitem{mueser:2017}
M.~M{\"u}ser, W.~Dapp, R.~Bugnicourt, P.~Sainsot, N.~Lesaffre, T.~Lubrecht,
  B.~Persson, K.~Harris, A.~Bennett, K.~Schulze, S.~Rohde, P.~Ifju, W.~Sawyer,
  T.~Angelini, H.~A. Esfahani, M.~Kadkhodaei, S.~Akbarzadeh, J.-J. Wu,
  G.~Vorlaufer, A.~Vernes, S.~Solhjoo, A.~I. Vakis, R.~L. Jackson, Y.~Xu,
  J.~Streator, A.~Rostami, D.~Dini, S.~Medina, G.~Carbone, F.~Bottiglione,
  L.~Afferrante, J.~Monti, L.~Pastewka, M.~O. Robbins, J.~A. Greenwood, Meeting
  the contact-mechanics challenge, Tribology Letters 65~(4) (2017) 118.
\newblock \href {https://doi.org/10.1007/s11249-017-0900-2}
  {\path{doi:10.1007/s11249-017-0900-2}}.

\bibitem{vakis:2018}
A.~Vakis, V.~Yastrebov, J.~Scheibert, L.~Nicola, D.~Dini, C.~Minfray,
  A.~Almqvist, M.~Paggi, S.~Lee, G.~Limbert, J.~Molinari, G.~Anciaux,
  S.~{Echeverri Restrepo}, A.~Papangelo, A.~Cammarata, P.~Nicolini,
  R.~Aghababaei, C.~Putignano, S.~Stupkiewicz, J.~Lengiewicz, G.~Costagliola,
  F.~Bosia, R.~Guarino, N.~Pugno, G.~Carbone, M.~Mueser, M.~Ciavarella,
  Modeling and simulation in tribology across scales: an overview, Tribology
  International (2018).
\newblock \href {https://doi.org/10.1016/j.triboint.2018.02.005}
  {\path{doi:10.1016/j.triboint.2018.02.005}}.

\bibitem{jacobs:2017}
T.~D.~B. Jacobs, A.~Martini, \href{https://doi.org/10.1115/1.4038130}{Measuring
  and {Understanding} {Contact} {Area} at the {Nanoscale}: {A} {Review}},
  Applied Mechanics Reviews 69~(6) (Nov. 2017).
\newblock \href {https://doi.org/10.1115/1.4038130}
  {\path{doi:10.1115/1.4038130}}.
\newline\urlprefix\url{https://doi.org/10.1115/1.4038130}

\bibitem{paggi:2020}
M.~Paggi, A.~Bemporad, J.~Reinoso,
  \href{https://doi.org/10.1007/978-3-030-20377-1_4}{Computational {Methods}
  for {Contact} {Problems} with {Roughness}}, in: M.~Paggi, D.~Hills (Eds.),
  Modeling and {Simulation} of {Tribological} {Problems} in {Technology},
  Springer International Publishing, Cham, 2020, pp. 131--178.
\newblock \href {https://doi.org/10.1007/978-3-030-20377-1_4}
  {\path{doi:10.1007/978-3-030-20377-1_4}}.
\newline\urlprefix\url{https://doi.org/10.1007/978-3-030-20377-1_4}

\bibitem{goryacheva:2021}
I.~G. Goryacheva, M.~Paggi, V.~L. Popov,
  \href{https://www.frontiersin.org/article/10.3389/fmech.2021.649792}{Editorial:
  Contact mechanics perspective of tribology}, Frontiers in Mechanical
  Engineering 7 (2021).
\newblock \href {https://doi.org/10.3389/fmech.2021.649792}
  {\path{doi:10.3389/fmech.2021.649792}}.
\newline\urlprefix\url{https://www.frontiersin.org/article/10.3389/fmech.2021.649792}

\bibitem{paggi:2022}
M.~Paggi, J.~Bonari, J.~Reinoso,
  \href{https://doi.org/10.1007/978-3-030-87312-7_37}{From the {Pioneering}
  {Contributions} by {Wriggers} to {Recent} {Advances} in {Computational}
  {Tribology}}, in: F.~Aldakheel, B.~Hudobivnik, M.~Soleimani, H.~Wessels,
  C.~Weißenfels, M.~Marino (Eds.), Current {Trends} and {Open} {Problems} in
  {Computational} {Mechanics}, Springer International Publishing, Cham, 2022,
  pp. 385--393.
\newblock \href {https://doi.org/10.1007/978-3-030-87312-7_37}
  {\path{doi:10.1007/978-3-030-87312-7_37}}.
\newline\urlprefix\url{https://doi.org/10.1007/978-3-030-87312-7_37}

\bibitem{bowden:1950}
F.~P. Bowden, D.~Tabor, The friction and lubrication of solids, Oxford:
  Clarendon Press, 1950.

\bibitem{archard:1957}
J.~Archard, Elastic deformation and the laws of friction, Proceedings of the
  Royal Society A 243~(1233) (1957) 190--205.
\newblock \href {https://doi.org/https://doi.org/10.1098/rspa.1957.0214}
  {\path{doi:https://doi.org/10.1098/rspa.1957.0214}}.

\bibitem{greenwood:1966}
J.~A. Greenwood, J.~B.~P. Williamson, Contact of nominally flat surfaces,
  Proceedings of the Royal Society of London A: Mathematical, Physical and
  Engineering Sciences 295 (1966) 300--319.
\newblock \href {https://doi.org/10.1098/rspa.1966.0242}
  {\path{doi:10.1098/rspa.1966.0242}}.

\bibitem{greenwood:1967}
J.~Greenwood, J.~Tripp, {The elastic contact of rough spheres}, Journal of
  Applied Mechanics, Transactions ASME 34~(1) (1967) 153--159.
\newblock \href {https://doi.org/10.1115/1.3607616}
  {\path{doi:10.1115/1.3607616}}.

\bibitem{bush:1975}
A.~Bush, R.~Gibson, T.~Thomas, The elastic contact of a rough surface, Wear
  35~(1) (1975) 87--111.
\newblock \href {https://doi.org/https://doi.org/10.1016/0043-1648(75)90145-3}
  {\path{doi:https://doi.org/10.1016/0043-1648(75)90145-3}}.

\bibitem{majumdar:1991}
A.~Majumdar, B.~Bhushan, Role of fractal geometry in roughness characterization
  and contact mechanics of surfaces, ASME Journal of Tribology 112 (1990)
  205--216.

\bibitem{borribrunetto:1999}
M.~Borri-Brunetto, A.~Carpinteri, B.~Chiaia, Scaling phenomena due to fractal
  contact in concrete and rock fractures, International Journal of Fracture 95
  (1999) 221--238.
\newblock \href {https://doi.org/10.1023/A:1018656403170}
  {\path{doi:10.1023/A:1018656403170}}.

\bibitem{persson:2001}
B.~Persson, Theory of rubber friction and contact mechanics, The Journal of
  Chemical Physics 115 (2001) 3840.
\newblock \href {https://doi.org/10.1063/1.1388626}
  {\path{doi:10.1063/1.1388626}}.

\bibitem{andersson:1981}
T.~Andersson, The boundary element method applied to two-dimensional contact
  problems with friction, in: Boundary Element Methods, Vol.~3, Springer,
  Berlin, Heidelberg, 1981, pp. 239--258.

\bibitem{johnson:1985}
K.~L. Johnson, Contact Mechanics, Cambridge University Press, Cambridge, UK,
  1985.

\bibitem{keer:1999}
I.~Polonsky, L.~Keer, A numerical method for solving rough contact problems
  based on the multi-level multi-summation and conjugate gradient techniques,
  Wear 231 (1999) 206--219.

\bibitem{bemporad:2015}
A.~Bemporad, M.~Paggi, Optimization algorithms for the solution of the
  frictionless normal contact between rough surfaces, International Journal of
  Solids and Structures 69--70 (2015) 94--105.

\bibitem{xu:2019}
Y.~Xu, R.~L. Jackson,
  \href{http://link.springer.com/10.1007/s40544-018-0229-3}{Boundary element
  method ({BEM}) applied to the rough surface contact vs. {BEM} in
  computational mechanics}, Friction 7~(4) (2019) 359--371.
\newblock \href {https://doi.org/10.1007/s40544-018-0229-3}
  {\path{doi:10.1007/s40544-018-0229-3}}.
\newline\urlprefix\url{http://link.springer.com/10.1007/s40544-018-0229-3}

\bibitem{paggi:2014}
M.~Paggi, R.~Pohrt, V.~L. Popov, {Partial-slip frictional response of rough
  surfaces}, Scientific Reports 4 (2014) 1--6.
\newblock \href {https://doi.org/10.1038/srep05178}
  {\path{doi:10.1038/srep05178}}.

\bibitem{pohrt:2014}
R.~Pohrt, Q.~Li, Complete boundary element formulation for normal and
  tangential contact problems, Physical Mesomechanics 17 (2014) 334--340.
\newblock \href {https://doi.org/10.1134/s1029959914040109}
  {\path{doi:10.1134/s1029959914040109}}.

\bibitem{vollebregt:2014}
E.~A. Vollebregt, \href{http://dx.doi.org/10.1016/j.jcp.2013.10.005}{{A new
  solver for the elastic normal contact problem using conjugate gradients,
  deflation, and an FFT-based preconditioner}}, Journal of Computational
  Physics 257~(PA) (2014) 333--351.
\newblock \href {https://doi.org/10.1016/j.jcp.2013.10.005}
  {\path{doi:10.1016/j.jcp.2013.10.005}}.
\newline\urlprefix\url{http://dx.doi.org/10.1016/j.jcp.2013.10.005}

\bibitem{anciaux:2010}
G.~Anciaux, J.~F. Molinari,
  \href{https://onlinelibrary.wiley.com/doi/10.1002/nme.2845}{Sliding of rough
  surfaces and energy dissipation with a {3D} multiscale approach: {SLIDING}
  {OF} {ROUGH} {SURFACES} {AND} {ENERGY} {DISSIPATION}}, International Journal
  for Numerical Methods in Engineering 83~(8-9) (2010) 1255--1271.
\newblock \href {https://doi.org/10.1002/nme.2845}
  {\path{doi:10.1002/nme.2845}}.
\newline\urlprefix\url{https://onlinelibrary.wiley.com/doi/10.1002/nme.2845}

\bibitem{conway:1966}
H.~D. Conway, S.~M. Vogel, K.~A. Farnham, S.~So, {Normal and shearing contact
  stresses in indented strips and slabs}, International Journal of Engineering
  Science 4~(4) (1966) 343--359.
\newblock \href {https://doi.org/10.1016/0020-7225(66)90036-x}
  {\path{doi:10.1016/0020-7225(66)90036-x}}.

\bibitem{bentall:1968}
R.~H. Bentall, K.~L. Johnson, {An elastic strip in plane rolling contact},
  International Journal of Mechanical Sciences 10~(8) (1968) 637--663.
\newblock \href {https://doi.org/10.1016/0020-7403(68)90070-2}
  {\path{doi:10.1016/0020-7403(68)90070-2}}.

\bibitem{greenwood:2012}
J.~Greenwood, J.~Barber, {Indentation of an elastic layer by a rigid cylinder},
  International Journal of Solids and Structures 49~(21) (2012) 2962--2977.

\bibitem{carbone:2013}
G.~Carbone, C.~Putignano, A novel methodology to predict sliding and rolling
  friction of viscoelastic materials: Theory and experiments, Journal of the
  Mechanics and Physics of Solids 61~(8) (2013) 1822--1834.
\newblock \href {https://doi.org/https://doi.org/10.1016/j.jmps.2013.03.005}
  {\path{doi:https://doi.org/10.1016/j.jmps.2013.03.005}}.

\bibitem{putignano:2014}
C.~Putignano, G.~Carbone, {A review of boundary elements methodologies for
  elastic and viscoelastic rough contact mechanics}, Physical Mesomechanics
  17~(4) (2014) 321--333.
\newblock \href {https://doi.org/10.1134/S1029959914040092}
  {\path{doi:10.1134/S1029959914040092}}.

\bibitem{putignano:2016}
C.~Putignano, G.~Carbone, D.~Dini,
  \href{http://arxiv.org/abs/1603.07598}{Theory of {Reciprocating} {Contact}
  for {Viscoelastic} {Solids}}, Physical Review E 93~(4) (2016) 043003, arXiv:
  1603.07598.
\newblock \href {https://doi.org/10.1103/PhysRevE.93.043003}
  {\path{doi:10.1103/PhysRevE.93.043003}}.
\newline\urlprefix\url{http://arxiv.org/abs/1603.07598}

\bibitem{carbone:2008}
G.~Carbone, L.~Mangialardi,
  \href{https://linkinghub.elsevier.com/retrieve/pii/S0022509607001147}{Analysis
  of the adhesive contact of confined layers by using a {Green}'s function
  approach}, Journal of the Mechanics and Physics of Solids 56~(2) (2008)
  684--706.
\newblock \href {https://doi.org/10.1016/j.jmps.2007.05.009}
  {\path{doi:10.1016/j.jmps.2007.05.009}}.
\newline\urlprefix\url{https://linkinghub.elsevier.com/retrieve/pii/S0022509607001147}

\bibitem{medina:2014}
S.~Medina, D.~Dini,
  \href{https://linkinghub.elsevier.com/retrieve/pii/S0020768314001395}{A
  numerical model for the deterministic analysis of adhesive rough contacts
  down to the nano-scale}, International Journal of Solids and Structures
  51~(14) (2014) 2620--2632.
\newblock \href {https://doi.org/10.1016/j.ijsolstr.2014.03.033}
  {\path{doi:10.1016/j.ijsolstr.2014.03.033}}.
\newline\urlprefix\url{https://linkinghub.elsevier.com/retrieve/pii/S0020768314001395}

\bibitem{pastewka:2014}
L.~Pastewka, M.~O. Robbins,
  \href{https://pnas.org/doi/full/10.1073/pnas.1320846111}{Contact between
  rough surfaces and a criterion for macroscopic adhesion}, Proceedings of the
  National Academy of Sciences 111~(9) (2014) 3298--3303.
\newblock \href {https://doi.org/10.1073/pnas.1320846111}
  {\path{doi:10.1073/pnas.1320846111}}.
\newline\urlprefix\url{https://pnas.org/doi/full/10.1073/pnas.1320846111}

\bibitem{popov:2017}
V.~L. Popov, R.~Pohrt, Q.~Li,
  \href{http://link.springer.com/10.1007/s40544-017-0177-3}{Strength of
  adhesive contacts: {Influence} of contact geometry and material gradients},
  Friction 5~(3) (2017) 308--325.
\newblock \href {https://doi.org/10.1007/s40544-017-0177-3}
  {\path{doi:10.1007/s40544-017-0177-3}}.
\newline\urlprefix\url{http://link.springer.com/10.1007/s40544-017-0177-3}

\bibitem{rey:2017}
V.~Rey, G.~Anciaux, J.-F. Molinari,
  \href{http://link.springer.com/10.1007/s00466-017-1392-5}{Normal adhesive
  contact on rough surfaces: efficient algorithm for {FFT}-based {BEM}
  resolution}, Computational Mechanics 60~(1) (2017) 69--81.
\newblock \href {https://doi.org/10.1007/s00466-017-1392-5}
  {\path{doi:10.1007/s00466-017-1392-5}}.
\newline\urlprefix\url{http://link.springer.com/10.1007/s00466-017-1392-5}

\bibitem{andersson:2011}
J.~Andersson, A.~Almqvist, R.~Larsson,
  \href{https://linkinghub.elsevier.com/retrieve/pii/S0043164811004583}{Numerical
  simulation of a wear experiment}, Wear 271~(11-12) (2011) 2947--2952.
\newblock \href {https://doi.org/10.1016/j.wear.2011.06.018}
  {\path{doi:10.1016/j.wear.2011.06.018}}.
\newline\urlprefix\url{https://linkinghub.elsevier.com/retrieve/pii/S0043164811004583}

\bibitem{brink:2021}
T.~Brink, L.~Frérot, J.-F. Molinari, \href{http://arxiv.org/abs/2004.00559}{A
  parameter-free mechanistic model of the adhesive wear process of rough
  surfaces in sliding contact}, Journal of the Mechanics and Physics of Solids
  147 (2021) 104238, arXiv: 2004.00559.
\newblock \href {https://doi.org/10.1016/j.jmps.2020.104238}
  {\path{doi:10.1016/j.jmps.2020.104238}}.
\newline\urlprefix\url{http://arxiv.org/abs/2004.00559}

\bibitem{frerot:2018}
L.~Frérot, R.~Aghababaei, J.-F. Molinari,
  \href{https://linkinghub.elsevier.com/retrieve/pii/S0022509617309742}{A
  mechanistic understanding of the wear coefficient: {From} single to multiple
  asperities contact}, Journal of the Mechanics and Physics of Solids 114
  (2018) 172--184.
\newblock \href {https://doi.org/10.1016/j.jmps.2018.02.015}
  {\path{doi:10.1016/j.jmps.2018.02.015}}.
\newline\urlprefix\url{https://linkinghub.elsevier.com/retrieve/pii/S0022509617309742}

\bibitem{mayeur:1995}
C.~Mayeur, P.~Sainsot, L.~Flamand, \href{https://doi.org/10.1115/1.2831270}{{A
  Numerical Elastoplastic Model for Rough Contact}}, Journal of Tribology
  117~(3) (1995) 422--429.
\newblock \href
  {http://arxiv.org/abs/https://asmedigitalcollection.asme.org/tribology/article-pdf/117/3/422/5938121/422\_1.pdf}
  {\path{arXiv:https://asmedigitalcollection.asme.org/tribology/article-pdf/117/3/422/5938121/422\_1.pdf}},
  \href {https://doi.org/10.1115/1.2831270} {\path{doi:10.1115/1.2831270}}.
\newline\urlprefix\url{https://doi.org/10.1115/1.2831270}

\bibitem{almqvist:2007}
A.~Almqvist, F.~Sahlin, R.~Larsson, S.~Glavatskih,
  \href{https://linkinghub.elsevier.com/retrieve/pii/S0301679X05003142}{On the
  dry elasto-plastic contact of nominally flat surfaces}, Tribology
  International 40~(4) (2007) 574--579.
\newblock \href {https://doi.org/10.1016/j.triboint.2005.11.008}
  {\path{doi:10.1016/j.triboint.2005.11.008}}.
\newline\urlprefix\url{https://linkinghub.elsevier.com/retrieve/pii/S0301679X05003142}

\bibitem{frerot:2019}
L.~Frérot, M.~Bonnet, J.-F. Molinari, G.~Anciaux,
  \href{https://linkinghub.elsevier.com/retrieve/pii/S0045782519302038}{A
  {Fourier}-accelerated volume integral method for elastoplastic contact},
  Computer Methods in Applied Mechanics and Engineering 351 (2019) 951--976.
\newblock \href {https://doi.org/10.1016/j.cma.2019.04.006}
  {\path{doi:10.1016/j.cma.2019.04.006}}.
\newline\urlprefix\url{https://linkinghub.elsevier.com/retrieve/pii/S0045782519302038}

\bibitem{frerot:2020}
L.~Frérot, G.~Anciaux, V.~Rey, S.~Pham-Ba, J.-F. Molinari,
  \href{https://joss.theoj.org/papers/10.21105/joss.02121}{Tamaas: a library
  for elastic-plastic contact of periodic rough surfaces}, Journal of Open
  Source Software 5~(51) (2020) 2121.
\newblock \href {https://doi.org/10.21105/joss.02121}
  {\path{doi:10.21105/joss.02121}}.
\newline\urlprefix\url{https://joss.theoj.org/papers/10.21105/joss.02121}

\bibitem{sahlin:2010}
F.~Sahlin, R.~Larsson, A.~Almqvist, P.~M. Lugt, P.~Marklund,
  \href{http://journals.sagepub.com/doi/10.1243/13506501JET658}{A mixed
  lubrication model incorporating measured surface topography. {Part} 1:
  {Theory} of flow factors}, Proceedings of the Institution of Mechanical
  Engineers, Part J: Journal of Engineering Tribology 224~(4) (2010) 335--351.
\newblock \href {https://doi.org/10.1243/13506501JET658}
  {\path{doi:10.1243/13506501JET658}}.
\newline\urlprefix\url{http://journals.sagepub.com/doi/10.1243/13506501JET658}

\bibitem{hyun:2004}
S.~Hyun, L.~Pei, J.-F. Molinari, M.~Robbins, Finite-element analysis of contact
  between elastic self-affine surfaces, Phys. Rev. E 70 (2004) 026117.
\newblock \href {https://doi.org/10.1103/PhysRevE.70.026117}
  {\path{doi:10.1103/PhysRevE.70.026117}}.

\bibitem{pei:2005}
L.~Pei, S.~Hyun, J.~Molinari, M.~O. Robbins, {Finite element modeling of
  elasto-plastic contact between rough surfaces}, Journal of the Mechanics and
  Physics of Solids 53~(11) (2005) 2385--2409.
\newblock \href {https://doi.org/10.1016/j.jmps.2005.06.008}
  {\path{doi:10.1016/j.jmps.2005.06.008}}.

\bibitem{bandeira:2004}
A.~A. Bandeira, P.~Wriggers, P.~{de Mattos Pimenta}, {Numerical derivation of
  contact mechanics interface laws using a finite approach for large 3D
  deformation}, International Journal for Numerical Methods in Engineering
  59~(2) (2004) 173--195.
\newblock \href {https://doi.org/10.1002/nme.867} {\path{doi:10.1002/nme.867}}.

\bibitem{yastrebov:2011}
V.~A. Yastrebov, J.~Durand, H.~Proudhon, G.~Cailletaud,
  \href{http://dx.doi.org/10.1016/j.crme.2011.05.006}{{Rough surface contact
  analysis by means of the Finite Element Method and of a new reduced model}},
  Comptes Rendus - Mecanique 339~(7-8) (2011) 473--490.
\newblock \href {https://doi.org/10.1016/j.crme.2011.05.006}
  {\path{doi:10.1016/j.crme.2011.05.006}}.
\newline\urlprefix\url{http://dx.doi.org/10.1016/j.crme.2011.05.006}

\bibitem{coutocarneiro:2020}
A.~M. {Couto Carneiro}, R.~{Pinto Carvalho}, F.~M. {Andrade Pires},
  \href{https://doi.org/10.1016/j.ijsolstr.2020.09.006}{{Representative contact
  element size determination for micromechanical contact analysis of
  self-affine topographies}}, International Journal of Solids and Structures
  206 (2020) 262--281.
\newblock \href {https://doi.org/10.1016/j.ijsolstr.2020.09.006}
  {\path{doi:10.1016/j.ijsolstr.2020.09.006}}.
\newline\urlprefix\url{https://doi.org/10.1016/j.ijsolstr.2020.09.006}

\bibitem{paggi:2018}
M.~Paggi, J.~Reinoso, A variational approach with embedded roughness for
  adhesive contact problems, Mechanics of Advanced Materials and Structures
  27~(20) (2020) 1731--1747.
\newblock \href {https://doi.org/10.1080/15376494.2018.1525454}
  {\path{doi:10.1080/15376494.2018.1525454}}.

\bibitem{ortiz:1999}
M.~Ortiz, A.~Pandolfi, Finite deformation irreversible cohesive elements for
  three-dimensional crack-propagation analysis, International Journal for
  Numerical Methods in Engineering 44 (1999) 1267--1282.
\newblock \href
  {https://doi.org/10.1002/(SICI)1097-0207(19990330)44:93.3.CO;2-Z}
  {\path{doi:10.1002/(SICI)1097-0207(19990330)44:93.3.CO;2-Z}}.

\bibitem{paggi:2016}
M.~Paggi, P.~Wriggers, \href{http://arxiv.org/abs/1604.05236}{Node-to-segment
  and node-to-surface interface finite elements for fracture mechanics},
  Computer Methods in Applied Mechanics and Engineering 300 (2016) 540--560,
  arXiv: 1604.05236.
\newblock \href {https://doi.org/10.1016/j.cma.2015.11.023}
  {\path{doi:10.1016/j.cma.2015.11.023}}.
\newline\urlprefix\url{http://arxiv.org/abs/1604.05236}

\bibitem{wriggers:2006}
P.~Wriggers, Computational Contact Mechanics, Springer-Verlag Berlin
  Heidelberg, 2006.
\newblock \href {https://doi.org/10.1007/978-3-540-32609-0}
  {\path{doi:10.1007/978-3-540-32609-0}}.

\bibitem{bonari:2021a}
J.~Bonari, M.~Paggi, J.~Reinoso,
  \href{https://linkinghub.elsevier.com/retrieve/pii/S0168874X21000895}{A
  framework for the analysis of fully coupled normal and tangential contact
  problems with complex interfaces}, Finite Elements in Analysis and Design 196
  (2021) 103605.
\newblock \href {https://doi.org/10.1016/j.finel.2021.103605}
  {\path{doi:10.1016/j.finel.2021.103605}}.
\newline\urlprefix\url{https://linkinghub.elsevier.com/retrieve/pii/S0168874X21000895}

\bibitem{bonari:2020}
J.~Bonari, M.~Paggi, Viscoelastic effects during tangential contact analyzed by
  a novel finite element approach with embedded interface profiles, Lubricants
  8~(12) (2020).
\newblock \href {https://doi.org/10.3390/lubricants8120107}
  {\path{doi:10.3390/lubricants8120107}}.

\bibitem{yu:2004}
N.~Yu, A.~A. Polycarpou,
  \href{https://linkinghub.elsevier.com/retrieve/pii/S0021979704005454}{Adhesive
  contact based on the {Lennard}–{Jones} potential: a correction to the value
  of the equilibrium distance as used in the potential}, Journal of Colloid and
  Interface Science 278~(2) (2004) 428--435.
\newblock \href {https://doi.org/10.1016/j.jcis.2004.06.029}
  {\path{doi:10.1016/j.jcis.2004.06.029}}.
\newline\urlprefix\url{https://linkinghub.elsevier.com/retrieve/pii/S0021979704005454}

\bibitem{sauer:2009}
R.~A. Sauer, P.~Wriggers,
  \href{https://linkinghub.elsevier.com/retrieve/pii/S0045782509002631}{Formulation
  and analysis of a three-dimensional finite element implementation for
  adhesive contact at the nanoscale}, Computer Methods in Applied Mechanics and
  Engineering 198~(49-52) (2009) 3871--3883.
\newblock \href {https://doi.org/10.1016/j.cma.2009.08.019}
  {\path{doi:10.1016/j.cma.2009.08.019}}.
\newline\urlprefix\url{https://linkinghub.elsevier.com/retrieve/pii/S0045782509002631}

\bibitem{mergel:2021}
J.~C. Mergel, J.~Scheibert, R.~A. Sauer,
  \href{https://linkinghub.elsevier.com/retrieve/pii/S002250962030418X}{Contact
  with coupled adhesion and friction: {Computational} framework, applications,
  and new insights}, Journal of the Mechanics and Physics of Solids 146 (2021)
  104194.
\newblock \href {https://doi.org/10.1016/j.jmps.2020.104194}
  {\path{doi:10.1016/j.jmps.2020.104194}}.
\newline\urlprefix\url{https://linkinghub.elsevier.com/retrieve/pii/S002250962030418X}

\bibitem{feeny:1994}
B.~Feeny, F.~C. Moon,
  \href{https://www.sciencedirect.com/science/article/pii/S0022460X84710650}{Chaos
  in a {Forced} {Dry}-{Friction} {Oscillator}: {Experiments} and {Numerical}
  {Modelling}}, Journal of Sound and Vibration 170~(3) (1994) 303--323.
\newblock \href {https://doi.org/https://doi.org/10.1006/jsvi.1994.1065}
  {\path{doi:https://doi.org/10.1006/jsvi.1994.1065}}.
\newline\urlprefix\url{https://www.sciencedirect.com/science/article/pii/S0022460X84710650}

\bibitem{simo:1992}
J.~Simo, T.~Laursen,
  \href{https://linkinghub.elsevier.com/retrieve/pii/004579499290540G}{An
  augmented lagrangian treatment of contact problems involving friction},
  Computers \& Structures 42~(1) (1992) 97--116.
\newblock \href {https://doi.org/10.1016/0045-7949(92)90540-G}
  {\path{doi:10.1016/0045-7949(92)90540-G}}.
\newline\urlprefix\url{https://linkinghub.elsevier.com/retrieve/pii/004579499290540G}

\bibitem{mostaghel:2005}
N.~Mostaghel,
  \href{https://linkinghub.elsevier.com/retrieve/pii/S0022460X04006030}{A
  non-standard analysis approach to systems involving friction}, Journal of
  Sound and Vibration 284~(3-5) (2005) 583--595.
\newblock \href {https://doi.org/10.1016/j.jsv.2004.06.041}
  {\path{doi:10.1016/j.jsv.2004.06.041}}.
\newline\urlprefix\url{https://linkinghub.elsevier.com/retrieve/pii/S0022460X04006030}

\bibitem{pennestri:2016}
E.~Pennestrì, V.~Rossi, P.~Salvini, P.~P. Valentini,
  \href{http://link.springer.com/10.1007/s11071-015-2485-3}{Review and
  comparison of dry friction force models}, Nonlinear Dynamics 83~(4) (2016)
  1785--1801.
\newblock \href {https://doi.org/10.1007/s11071-015-2485-3}
  {\path{doi:10.1007/s11071-015-2485-3}}.
\newline\urlprefix\url{http://link.springer.com/10.1007/s11071-015-2485-3}

\bibitem{vigue:2017}
P.~Vigué, C.~Vergez, S.~Karkar, B.~Cochelin,
  \href{https://linkinghub.elsevier.com/retrieve/pii/S0022460X16306277}{Regularized
  friction and continuation: {Comparison} with {Coulomb}'s law}, Journal of
  Sound and Vibration 389 (2017) 350--363.
\newblock \href {https://doi.org/10.1016/j.jsv.2016.11.002}
  {\path{doi:10.1016/j.jsv.2016.11.002}}.
\newline\urlprefix\url{https://linkinghub.elsevier.com/retrieve/pii/S0022460X16306277}

\bibitem{bonari:2021}
J.~Bonari, M.~Paggi, J.~Reinoso,
  \href{https://www.sciencedirect.com/science/article/pii/S0168874X21000895}{A
  framework for the analysis of fully coupled normal and tangential contact
  problems with complex interfaces}, Finite Elements in Analysis and Design 196
  (2021) 103605.
\newblock \href {https://doi.org/https://doi.org/10.1016/j.finel.2021.103605}
  {\path{doi:https://doi.org/10.1016/j.finel.2021.103605}}.
\newline\urlprefix\url{https://www.sciencedirect.com/science/article/pii/S0168874X21000895}

\bibitem{mandelbrot:1977}
B.~B. Mandelbrot, Fractal Geometry of Nature, San Francisco: W.H. Freeman,
  1977.

\bibitem{fractals:1988}
M.~F. Barnsley, R.~L. Devaney, B.~B. Mandelbrot, H.-O. Peitgen, The Science of
  Fractal Images, Springer-Verlag New York, 1988.

\bibitem{perez:2019}
F.~Pérez-Ràfols, A.~Almqvist,
  \href{https://linkinghub.elsevier.com/retrieve/pii/S0301679X18305607}{Generating
  randomly rough surfaces with given height probability distribution and power
  spectrum}, Tribology International 131 (2019) 591--604.
\newblock \href {https://doi.org/10.1016/j.triboint.2018.11.020}
  {\path{doi:10.1016/j.triboint.2018.11.020}}.
\newline\urlprefix\url{https://linkinghub.elsevier.com/retrieve/pii/S0301679X18305607}

\bibitem{hills:1994}
D.~Hills,
  \href{https://www.sciencedirect.com/science/article/pii/0043164894901732}{Mechanics
  of fretting fatigue}, Wear 175~(1) (1994) 107--113.
\newblock \href {https://doi.org/https://doi.org/10.1016/0043-1648(94)90173-2}
  {\path{doi:https://doi.org/10.1016/0043-1648(94)90173-2}}.
\newline\urlprefix\url{https://www.sciencedirect.com/science/article/pii/0043164894901732}

\bibitem{nowell:2006}
D.~Nowell, D.~Dini, D.~A. Hills, Recent developments in the understanding of
  fretting fatigue, Engineering Fracture Mechanics 73~(2) (2006) 207--222.
\newblock \href
  {https://doi.org/https://doi.org/10.1016/j.engfracmech.2005.01.013}
  {\path{doi:https://doi.org/10.1016/j.engfracmech.2005.01.013}}.

\end{thebibliography}

\end{document}